\documentclass{aa}  
\usepackage{graphicx}

\usepackage{txfonts}
\usepackage{hyperref}
\newcommand\tiu{J\,m$^{-2}$K$^{-1}$s$^{-1/2}$}
\newcommand\gcc{g\,cm$^{-3}$}
\def\gg{G!kún$\parallel$'hòmdímà}
\def\ggg{G!ò'é!Hú}
\newcommand\psj{\ref@jnl{PSJ}}

\begin{document}

   \title{The visible and thermal light curve of the large Kuiper belt object (50000)~Quaoar\thanks{This paper includes data obtained with the Herschel Space Observatory: {\it Herschel} is an ESA space observatory with science instruments provided by European-led Principal Investigator consortia and with important participation from NASA. } \thanks{Tables 1 and 3 are only available in electronic form at the CDS via anonymous ftp to cdsarc.u-strasbg.fr (130.79.128.5) or via http://cdsweb.u-strasbg.fr/cgi-bin/qcat?J/A+A/}}

   \author{
          C.~Kiss\inst{1,2,3}
          \and
          T.G.~M\"uller\inst{4}
          \and
          G.~Marton\inst{1,2}
          \and
          R.~Szak\'ats\inst{1,2}
          \and
          A.~P\'al\inst{1,2}
          \and
          L.~Molnár\inst{1,2,3}
          \and 
          E.~Vilenius\inst{5}
          \and
          M.~Rengel\inst{5}
          \and 
          J.L.~Ortiz\inst{6} 
          \and
          E.~Fern\'andez-Valenzuela\inst{7}
          }

   \institute{
            Konkoly Observatory, HUN-REN Research Centre for Astronomy and Earth Sciences, Konkoly Thege 15-17, H-1121 Budapest, Hungary
         \and
           CSFK, MTA Centre of Excellence, Budapest, Konkoly Thege 15-17, H-1121, Budapest, Hungary
         \and
             ELTE Eötvös Loránd University, Institute of Physics and Astronomy, Budapest, Hungary
         \and
         Max-Planck-Institut für extraterrestrische Physik, Garching, Germany
         \and
         Max-Planck-Institut für Sonnensystemforschung, Justus-von-Liebig-Weg 3, 37077 Göttingen, Germany
         \and 
         Instituto de Astrof\'\i{}sica de Andaluc\'\i{}a, IAA-CSIC, Glorieta de la Astronom\'\i{}a
             s/n, E-18008 Granada, Spain
        \and
        Florida Space Institute, UCF, 12354 Research Parkway, Partnership 1 building, Room 211, Orlando, USA
             }

   \date{Received \today; accepted \today}

  \abstract{
 Recent stellar occultations have allowed accurate instantaneous size and apparent shape determinations of the large Kuiper belt object (50000)~Quaoar and the detection of  two rings with spatially variable optical depths.
  In this paper we present new visible range light curve data of Quaoar from the Kepler/K2 mission, and thermal light curves at 100 and 160\,$\mu$m obtained with Herschel/PACS. The K2 data provide a single-peaked period of 8.88\,h, very close to the previously determined 8.84\,h, and it favours an asymmetric double-peaked light curve with a 17.76\,h period. We clearly detected a thermal light curve with relative amplitudes of $\sim$10\%  at 100 and at 160\,$\mu$m.
  A detailed thermophysical  modelling of the system shows that the measurements can be best fit 
  with a triaxial ellipsoid shape,  a volume-equivalent diameter of 1090\,km, and axis ratios of a/b\,=\,1.19 and b/c\,=\,1.16. { This shape matches the published occultation shape}, as well as visual and thermal light curve data. The radiometric size uncertainty remains relatively large ($\pm$40\,km) as the ring and satellite contributions to the system-integrated flux densities are unknown.  In the less likely case of negligible ring or satellite contributions, Quaoar would have a size above 1100\,km and a thermal inertia $\leq$\,10\,\tiu. 
  A large and dark Weywot in combination with a possible ring contribution would lead to a size below 1080\,km in combination with a thermal inertia $\gtrsim$\,10\,\tiu, notably higher than that of smaller Kuiper belt objects with
   similar albedo and colours. We find that Quaoar's density is in the range 1.67-1.77\,\gcc, significantly lower than previous estimates. This density value closely matches the relationship observed between the size and density of the largest Kuiper belt objects.
  }


   \maketitle
%

\section{Introduction}

The large Kuiper belt object (50000)~Quaoar is on a hot classical orbit \citep{Gladman2008} with a semi-major axis of 43.7\,au and an inclination of 8\fdg0. Its spectral type is moderately red (IR) and its geometric albedo is $\sim$0.12 \citep{Pereira2023}. Quaoar is one of the key bodies to understand planetesimal formation in the early  Solar System: this is one of the outer Solar System objects with a large main body ($\sim$1000\,km)  and a small satellite with a mass ratio q\,$\lesssim$\,0.1, in contrast to the typically nearly equal sized binaries of smaller sizes \citep{Noll2020,Holler2021}. With its size between that of the dwarf planet Sedna and Pluto's moon Charon, Quaoar was assumed to be a transitional object between volatile-poor and volatile-rich objects \citep{Brown2011}.
{ Recent spectra obtained with the James Webb Space Telescope \citep{Emery2023} show the presence of the light hydrocarbons methane { and  ethane, and } broad absorptions between 2.7 and 3.6\,$\mu$m indicative of complex organic molecules. These are interpreted as the products of irradiation of methane, suggesting a resupply of methane to the surface.} Crystalline water ice has also been reported \citep{Jewitt2004,Emery2023}, implying that the temperature has been as high as 110\,K at some time during the last 10\,Myr, indicating impact exposure or cryovolcanism on Quaoar.

The physical properties of Quaoar (e.g. shape, size, density, rotation)  are very important in constraining formation and evolution theories. Recent stellar occultations have revealed a complex ring system around Quaoar \citep{Braga-Ribas2013,Morgado2023,Pereira2023}, and have also provided accurate determinations of the apparent size and shape of Quaoar's main body at the time of the occultation events.
{ Using the 2022 occultation data, Quaoar's apparent limb could be fitted with an ellipse with semi-major axis of a'\,=\,579$\pm$4\,km and apparent oblateness of $\epsilon$'\,=\,0.12. Two inhomogeneous rings have been discovered, Q1R at a distance of 4157\,km and Q2R at 2520\,km, both outside Quaoar's classical Roche limit.}
While { multi-chord} stellar occultations are the most precise ground-based methods to measure apparent size and shape, understanding the object is not possible without other important physical properties such as rotational characteristics (period, light curve shape, and spin axis orientation), surface temperature, thermal inertia, and chemical composition \citep{Ortiz2020}. There have been a relatively large number of thermal emission measurements for 178 Centaurs and trans-Neptunian objects (TNOs)
measured with the Spitzer Space Telescope, the Herschel Space Observatory, and the WISE space telescope \citep{Stansberry2008,Muller2018,Muller2020}. While these measurements offer invaluable insights into the thermal properties and surface characteristics of these targets \citep[see e.g.][]{Lellouch2013,Lacerda2014}, most of these measurements just provided a snapshot of the thermal emission of these targets and were used as representative flux densities, not considering any dependence on the rotational phase.  Thermal emission light curves that would allow a more thorough characterisation are available only for a handful of TNOs. For example, \citet{Lellouch2017} obtained evidence of emissivity effects in the Pluto-Charon system using multi-band thermal emission light curves observed by the Spitzer and Herschel space telescopes. \citet{Lockwood2014} obtained Spitzer/MIPS 70\,$\mu$m light curve data of Haumea, and \citet{SantsSanz2017} analysed the Herschel/PACS light curves of Haumea, 2003\,VS$_2$ and 2003\,AZ$_{84}$, and obtained low surface thermal intertias for these targets. The thermal emission light curve of the dwarf planet Haumea was re-evaluated by \citet{Muller2019} considering occultation data, and the analysis of these data were able to  constrain the main thermal characteristics of Haumea's surface as well as the contribution of Haumea's ring and satellites to the thermal emission. The recent discovery of the rings and its previously known satellite make Quaoar a similarly complex system. 

In this paper we present visible range light curve measurements using data from the K2 mission of the Kepler Space Telescope and thermal emission light curves at 100\,$\mu$m and 160\,$\mu$m observed with the PACS camera of the Herschel Space Observatory.
We estimate the potential thermal emission contributions of Quaoar's rings and its satellite, Weywot. 
These data are used to obtain a thermal emission model of Quaoar, constraining its shape, size, density, and surface thermal properties. 

\section{Kepler-K2 light curve \label{sect:k2}}
\label{sec:k2}

In Campaign 9 of the K2 mission \citep{K2,Kiss2020} the Kepler Space Telescope looked in the direction of the Galactic bulge and Quaoar was moving through a dense stellar field, making the photometry especially challenging, and a significant part of the data had to be discarded.
{ This initial selection resulted in 601 light curve samples covering} a period of $\sim$26\,days. Data reduction and photometry were performed in the same way as in \citet{K23}, and we refer to that paper for the details. The K2 light curve data of Quaoar is presented in Table~\ref{tab:k2q}. 
\begin{table*}
    \begin{center}
     \caption{K2 light curve data of Quaoar. }
    \begin{tabular}{cccccc}
    \hline
    JD$_{mean}$ & m$_{K2}$ & $\delta$m$_{K2}$ & r$_h$ & $\Delta$ & $\alpha$\\
     & (mag) & (mag) & (au) & (au) & (deg) \\
    \hline
    2457501.097015 &    18.781 &    0.130 & 42.9588     & 43.1714   &   1.3057\\
    2457501.117449 &    19.095 &    0.106 & 42.9588  &   43.1710   &   1.3058\\
    2457501.137883 &    19.212 &    0.072 & 42.9588   &  43.1707   &   1.3059\\
    2457501.158316 &    19.434 &    0.026 & 42.9588   &  43.1703   &   1.3060\\
    2457501.240051 &    19.406 &    0.235 & 42.9588   &  43.1690   &  1.3065\\ 
    \hline
    \end{tabular}
    \label{tab:k2q}
    \end{center}
    { Note.} {The columns are (1) spacecraft-centric, mean Julian date (t$_{exp}$\,=\,29.4\,s); (2) K2-band brightness (mag); (3) brightness uncertainty (mag); (4) heliocentric distance; (5) observer distance; (6) phase angle. The  `gaps' in the data are due to excluded data points.
     Only the first five rows are shown, the table is available in its entirety in electronic format.}
\end{table*}

\begin{figure}
   \centering
   \includegraphics[width=\columnwidth]{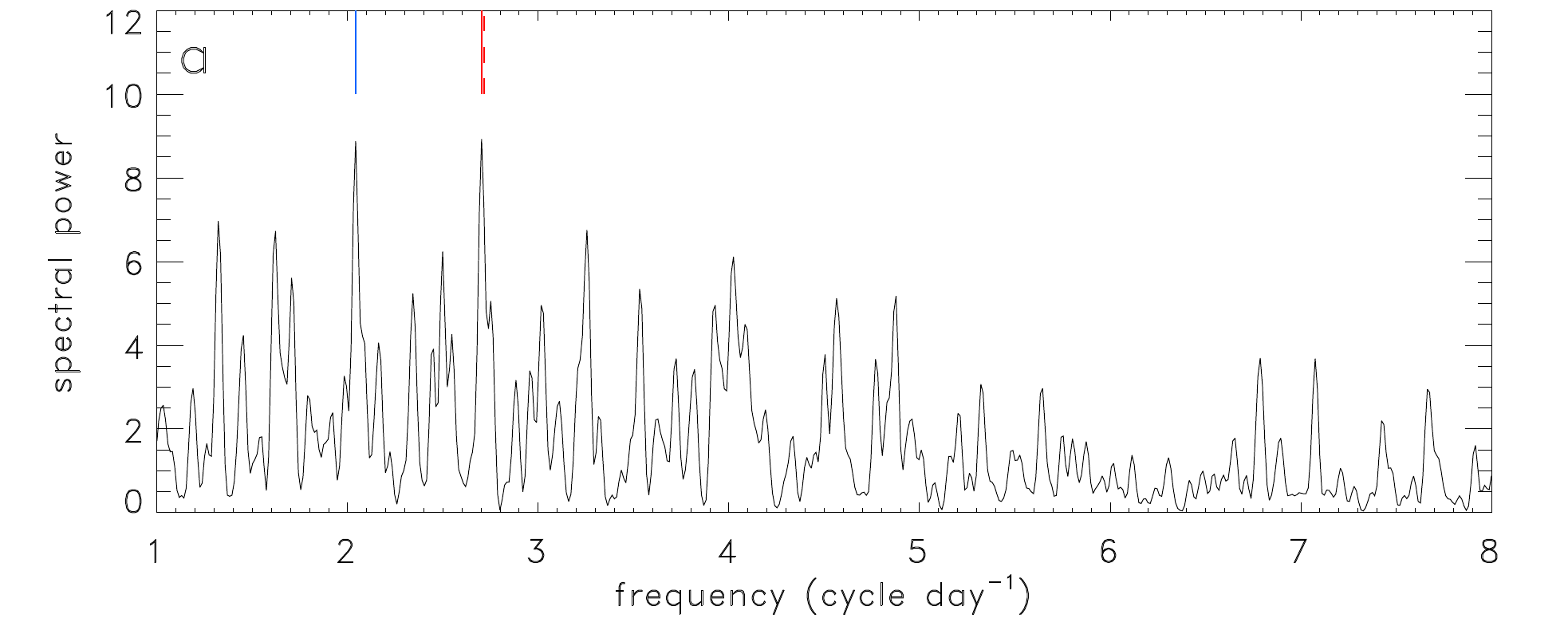}

     \includegraphics[width=\columnwidth]{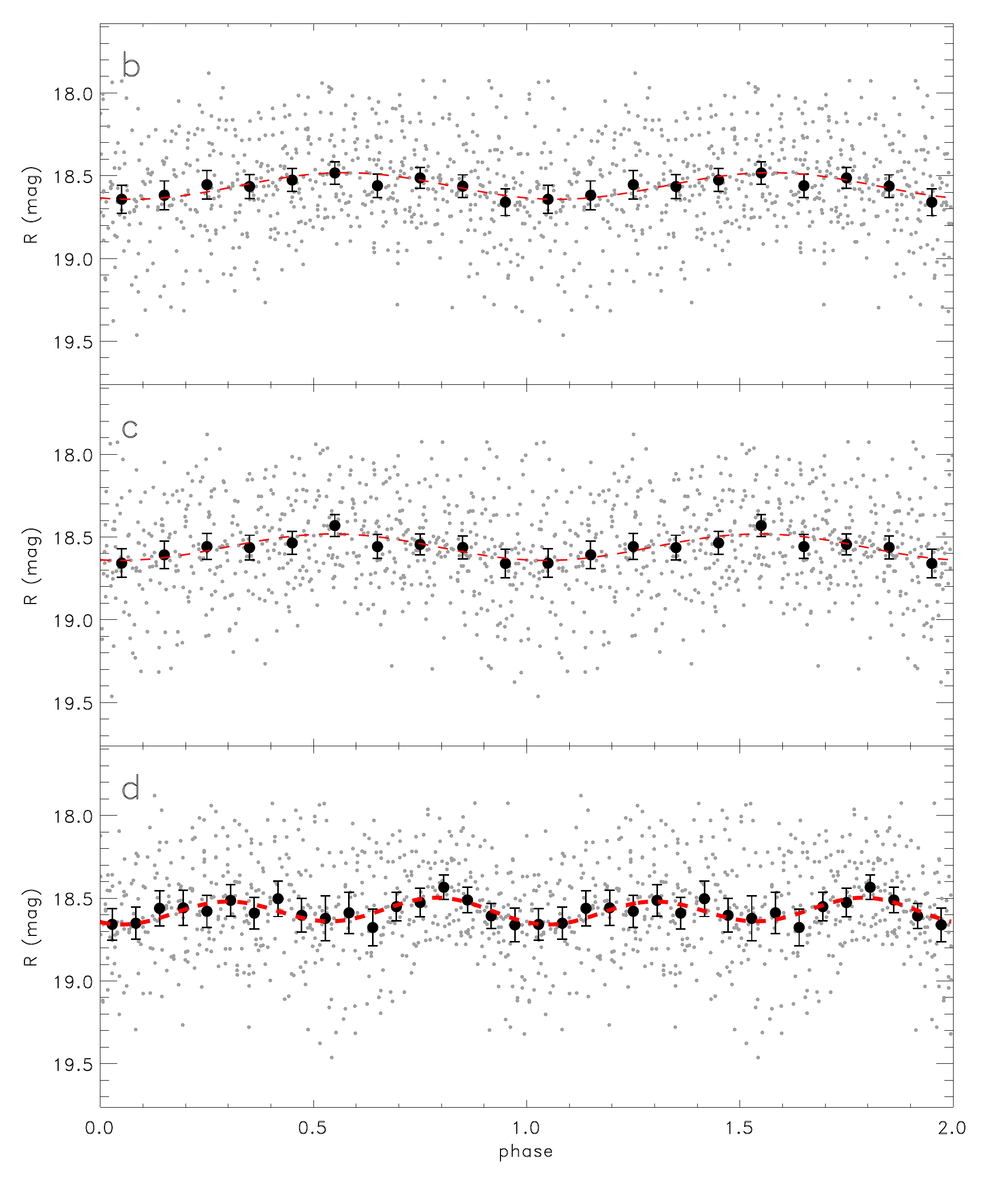}
      \caption{{ Kepler-K2 light curve data of Quaoar.}
      (a) Lomb--Scargle periodogram of the Kepler-K2 light curve of Quaoar. The red marks represent the frequencies that correspond to the previously identified period of P\,=\,8.8402\,h, and the new period P\,=\,8.876\,h. { The blue mark corresponds to the period of 11.748\,h;} (b) Kepler-K2 light curve folded with the original P\,=\,8.84\,h period. The red curve is a best-fit sine function using this period; 
      (c) Same as (b), but folded with the P\,=\,8.884\,h period; (d) Same as (b), but folded with the 17.752\,h(=\,2$\times$8.876\,h) period.
              }
         \label{fig:tesslc}
   \end{figure}

The light curves obtained were analysed with a residual minimization algorithm \citep{Pal2016,Molnar2018}. 
As demonstrated in \citet{Molnar2018}, the best-fit  frequencies obtained with this method are identical to the results of  Lomb–Scargle periodogram and fast Fourier transform analyses,  typically with notably smaller general uncertainty in the residuals. In the case of our Quaoar light curve these methods provided identical results. The frequency range f\,$\leq$\,1\,cycle/day (c/d) is strongly noise dominated \citep[see a more detailed discussion in][]{K23}, and therefore we searched the frequency range of f\,=\,1-10\,c/d for characteristic frequencies. In Fig.~\ref{fig:tesslc}a we present the Lomb--Scargle periodogram, but the residual  minimisation method resulted in an almost identical residual spectrum, with the same main peaks. 
The strongest frequency was identified at f\,=\,2.704$\pm$0.022\,c/d and a corresponding period of P\,=\,8.876$\pm$0.072\,h, which is just slightly different from the P\,=\,8.84\,h reported by \citet{Ortiz2003quaoar} and \citet{Thirouin2010}, and compatible with those values within the uncertainties. A single-peaked light curve is expected from an oblate spheroid with surface albedo variegations, a shape typically expected from large planetary bodies in hydrostatic equilibrium.

%

The Lomb--Scargle periodogram also shows another peak with very similar strength at P\,=\,11.748$\pm$0.109\,h, which has not been reported in any other dataset. While the source of this second peak could not be identified, as the 8.84\,h period was reported in previous papers, we consider our 8.88\,h period as a confirmation of the single-peaked rotation period of Quaoar reported earlier. 

A double-peaked light curve is a strong indication of a non-axisymmetric body, usually considered as a triaxial ellipsoid, where the brightness variations are caused by the changing apparent  cross-section due to rotation; previous light curve studies of Quaoar suggested that its light curve may be double-peaked \citep{Ortiz2003quaoar,Thirouin2010}. We calculated the significance of a double-peaked light curve over the single-peaked curve using the double period of 2$\times$8.876\,h\,=\,17.752\,h { first}, following \citet{Pal2016}. We compared the uncertainty-weighted differences between the corresponding bins in the first and second halfs of the folded light curve (see also Fig.~\ref{fig:tesslc}d), providing a significance of S\,$\approx$\,2.7, which corresponds to a $\sim$99\% probability that the two half periods of the double-peaked light curve are different.
{ A similar test can be performed using a Student's t-test following  \citet{Hromakina2019}, among others. Both calculations indicate that the double-peaked light curve is preferred with $\sim$98\% confidence, with a slight dependence on the number of bins chosen.}
The full amplitudes of the two half-periods of the  double-peaked light curve are estimated to be $\Delta m_1$\,=\,0.12$\pm$0.02\,mag, and $\Delta m_2$\,=\,0.16$\pm$0.02\,mag. 

The single- and double-peaked light curve   have very similar peak-to-peak amplitudes with A$_s$\,=\,0.161$\pm$0.046\,mag and A$_d$\,=\,0.161$\pm$0.056\,mag, considering the maximum and minimum binned values and standard error propagation in the calculation of the uncertainties \citep[see also][]{K23}. These are compatible with the amplitudes 0.133$\pm$0.028\,mag and 0.15$\pm$0.04\,mag, obtained by \citet{Ortiz2003quaoar} and \citet{Thirouin2010} considering the error bars.

\section{Far-infrared thermal light curve with Herschel/PACS \label{sect:pacslc}}

Far-infrared thermal light curve measurements were performed with the Photometer Array Camera and Spectrometer (PACS) camera on board the Herschel Space Observatory \citep{PACS}, in the framework of the Open Time Proposal `OT1\_evileniu\_1: Probing the extremes of the outer Solar System: Short-term variability of the largest, the densest, and the most distant TNOs from PACS photometry' \citep{evileniu1}. Observations at three epochs with equal length (82 repetitions each) and partial background overlap were executed, using the PACS 100/160\,$\mu$m filter combination. The main characteristics of the observations are listed in Table~\ref{table:lc1}. We note the mismatch between the sequence of the measurements and the OBSIDs that was caused by a later re-scheduling of previously failed observations. The times of the measurements were set in a way that first, it allowed Quaoar to move enough between the measurement blocks so that they could  be used for mutual background subtraction, and second, the three observations provided a full phase coverage assuming the double-peaked 17.7\,h rotation period without overlap.  While this set-up was optimal in the sense that it provides a maximum rotational phase coverage with the minimum observing time required, it does not allow cross-calibration of flux densities of the same rotational phases from different epochs. 


\begin{table*}[!ht]
\caption[]{Herschel/PACS photometer scan map observations of Quaoar. }
\small
\begin{tabular}{rlrlrcrccccc}
\hline
OD & OBSID &  SAA & UTC Start time & Dur. & Band & Rep. 
& r$_h$ & $\Delta$ & $\alpha$ &
$\overline{F}_{100}$ & $\overline{F}_{160}$\\
  & & (deg) & & (min) & & & (au) & (au) & (\degr) &
  (mJy) & (mJy)\\
\hline 
 871 & 1342229967 &  17.7 & 2011 Oct 01 22:15:48 & 393 & 100/160  & 82 
 & 43.114 & 43.408 & 1.28  
 & 39.49$\pm$0.47 & 29.84$\pm$0.68 \\
 872 & 1342230111 &   18.7 & 2011 Oct 02 22:06:35 & 396 & 100/160 & 82 & 43.114 & 43.424 & 1.27   & 37.53$\pm$0.36 & 28.33$\pm$0.77 \\
 873 & 1342230064 &   19.6 & 2011 Oct 03 21:39:02 & 397 & 100/160 & 82 &  43.114 & 43.439 & 1.27   & 41.04$\pm$0.42 & 33.50$\pm$0.55  \\
\hline 
\end{tabular}
\label{table:lc1}                       
{ Note.} {All data were taken in `mini-scan map' observing mode, using high gain and satellite scan speed of 20$^{\prime\prime}$\,s$^{-1}$. The columns are: OD,  operational day; OBSID,  Herschel observation ID of the measurement; 
SAA, solar aspect angle; Dur., duration of observation in minutes; Band, filter--band combination; Rep.,  repetition of entire scan map; r$_h$, heliocentric distance; $\Delta$, observer distance; $\alpha$, phase angle;
$\overline{F}_{100}$ and $\overline{F}_{160}$, mean in-band flux density in the 100 and 160\,$\mu$m bands of the measurement block. All measurements were performed with 3.0$^{\prime}$ scan-leg length, ten scan legs,  4.0$^{\prime\prime}$ scan-leg separation, and 110\degr\, satellite scan angle with respect to instrument reference frame. } 
\end{table*}

\begin{figure}[!ht]
    \centering
    \includegraphics[width=\columnwidth]{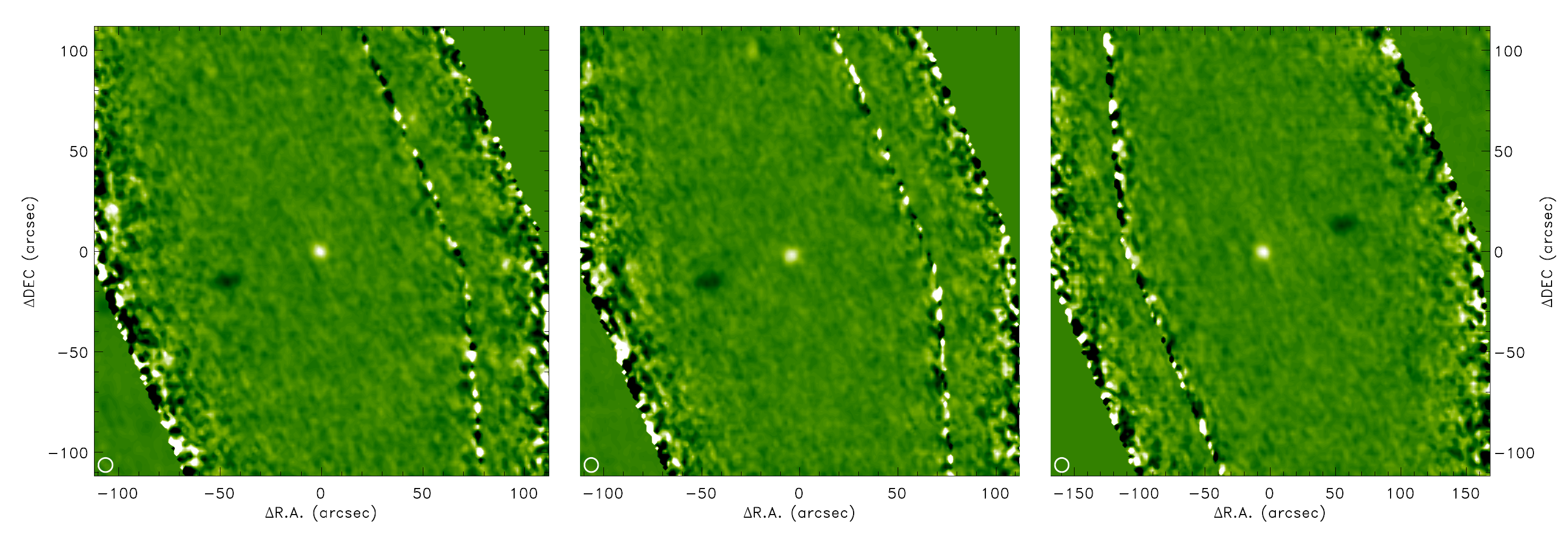}
    \includegraphics[width=\columnwidth]{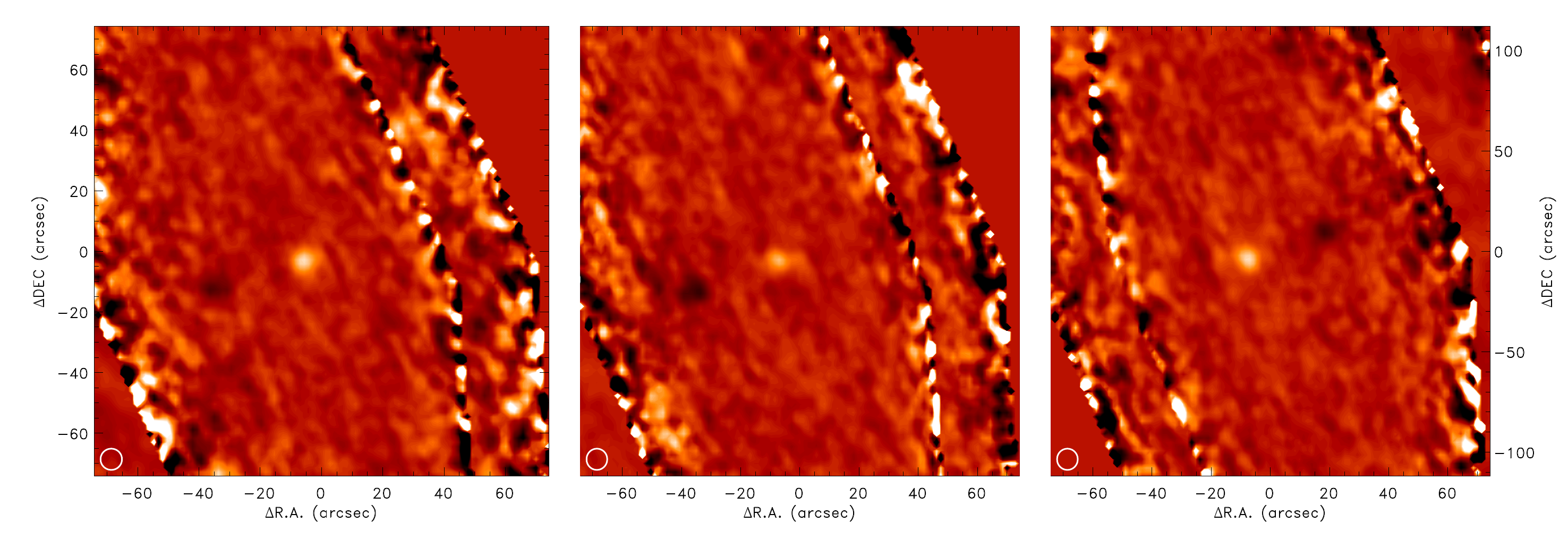}
    \caption{Example images of the Herschel/PACS light curve data. The 100\,$\mu$m (top row) and 160\,$\mu$m (bottom row) LCDIFF images with the OBSIDs and starting repetitions 134229967/001, 13422230111/001, and 13422230111/076 are shown. The photometry was performed on the bright source at the centre  of the image. The dark spot in each panel is the image of Quaoar on the respective background image. In all images the colours are drawn from -3$\sigma$ (black) and +3$\sigma$ (white). The black and white stripes are the edges of the single scan maps. The coverage is significantly lower, and therefore the noise is higher towards the edges than in the centre.} 
    \label{fig:lcdiff}
\end{figure}

\begin{figure*}
    \centering
    \hbox{\includegraphics[width=0.5\textwidth]{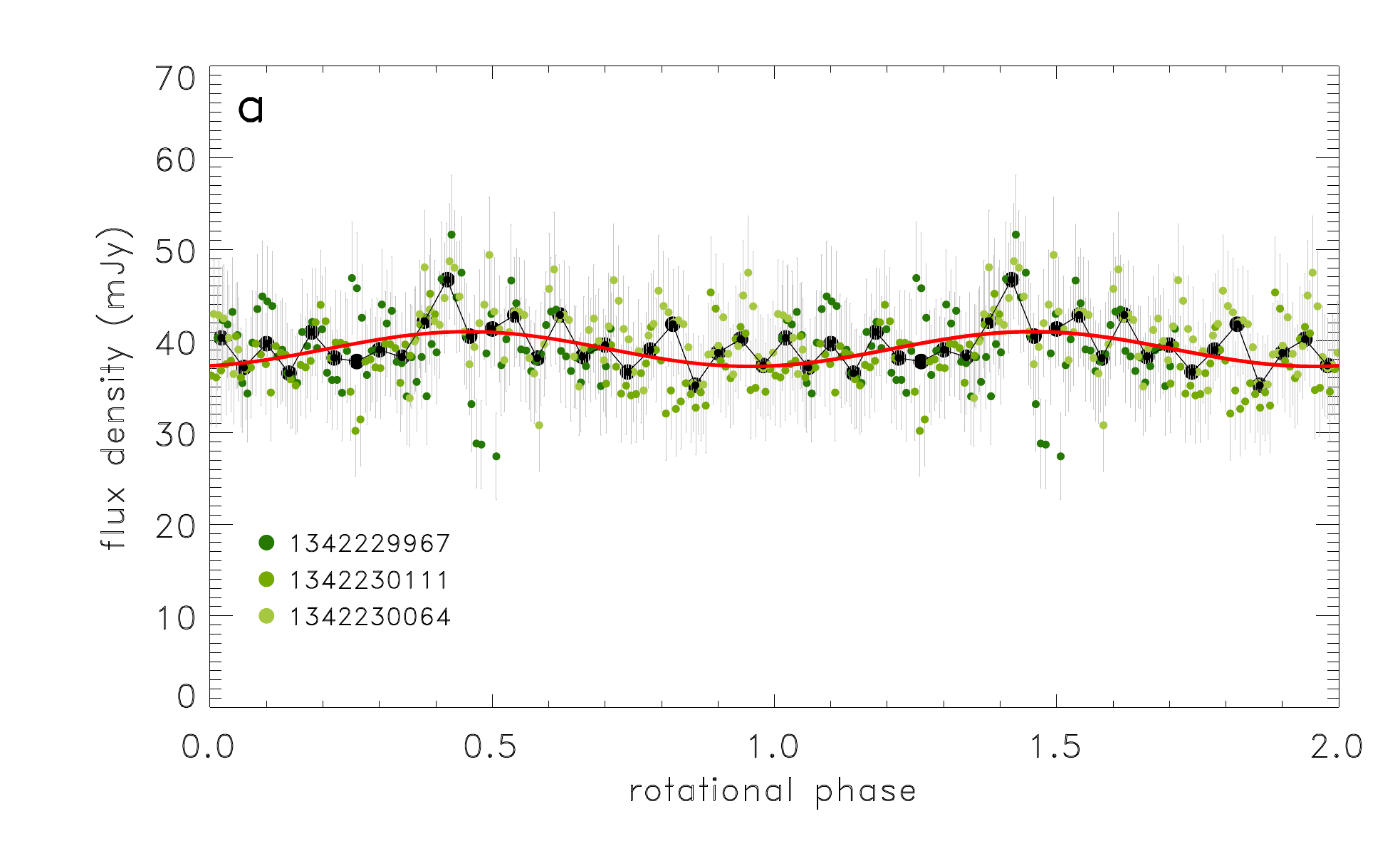}
            \includegraphics[width=0.5\textwidth]{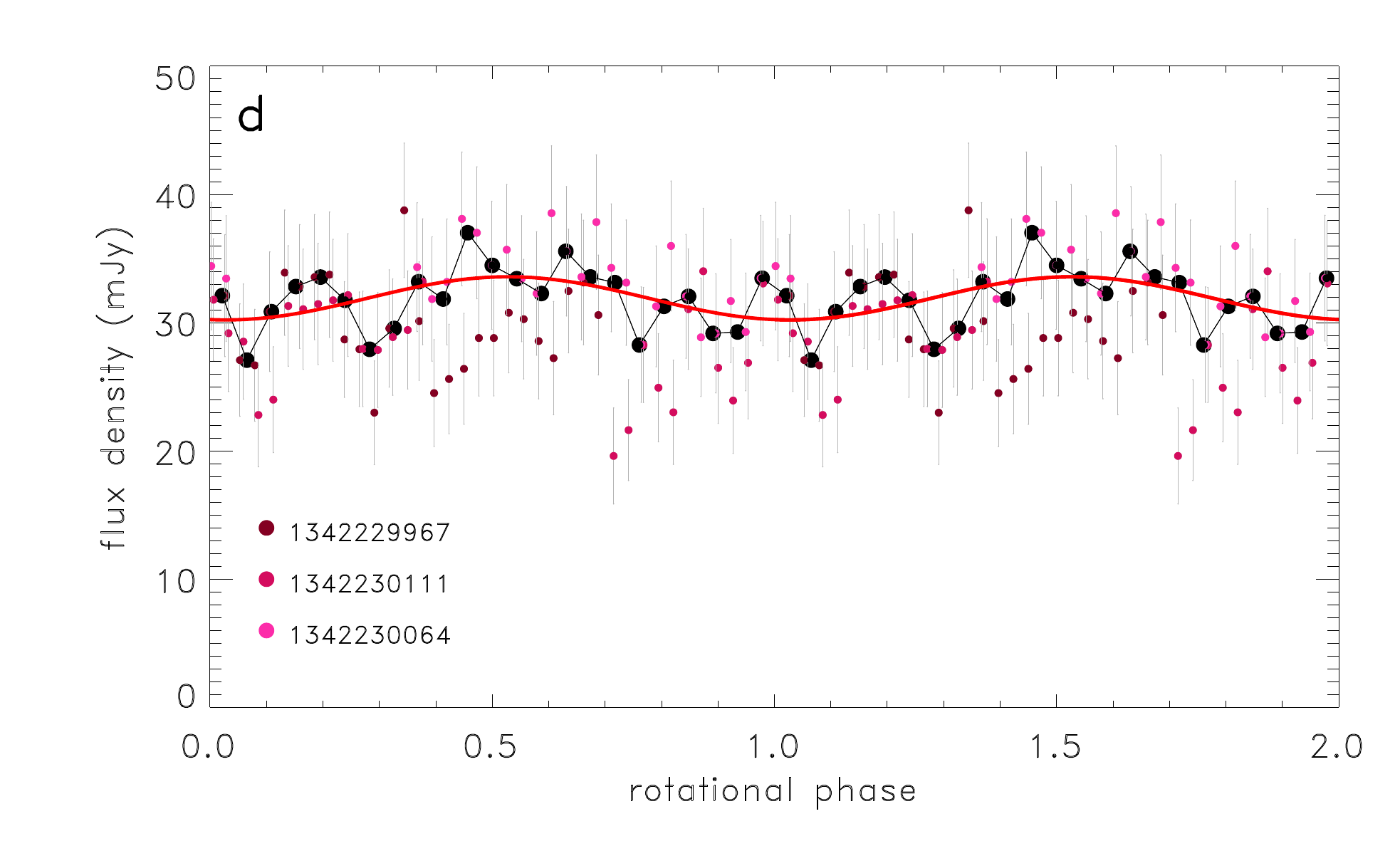}}
    \hbox{\includegraphics[width=0.5\textwidth]{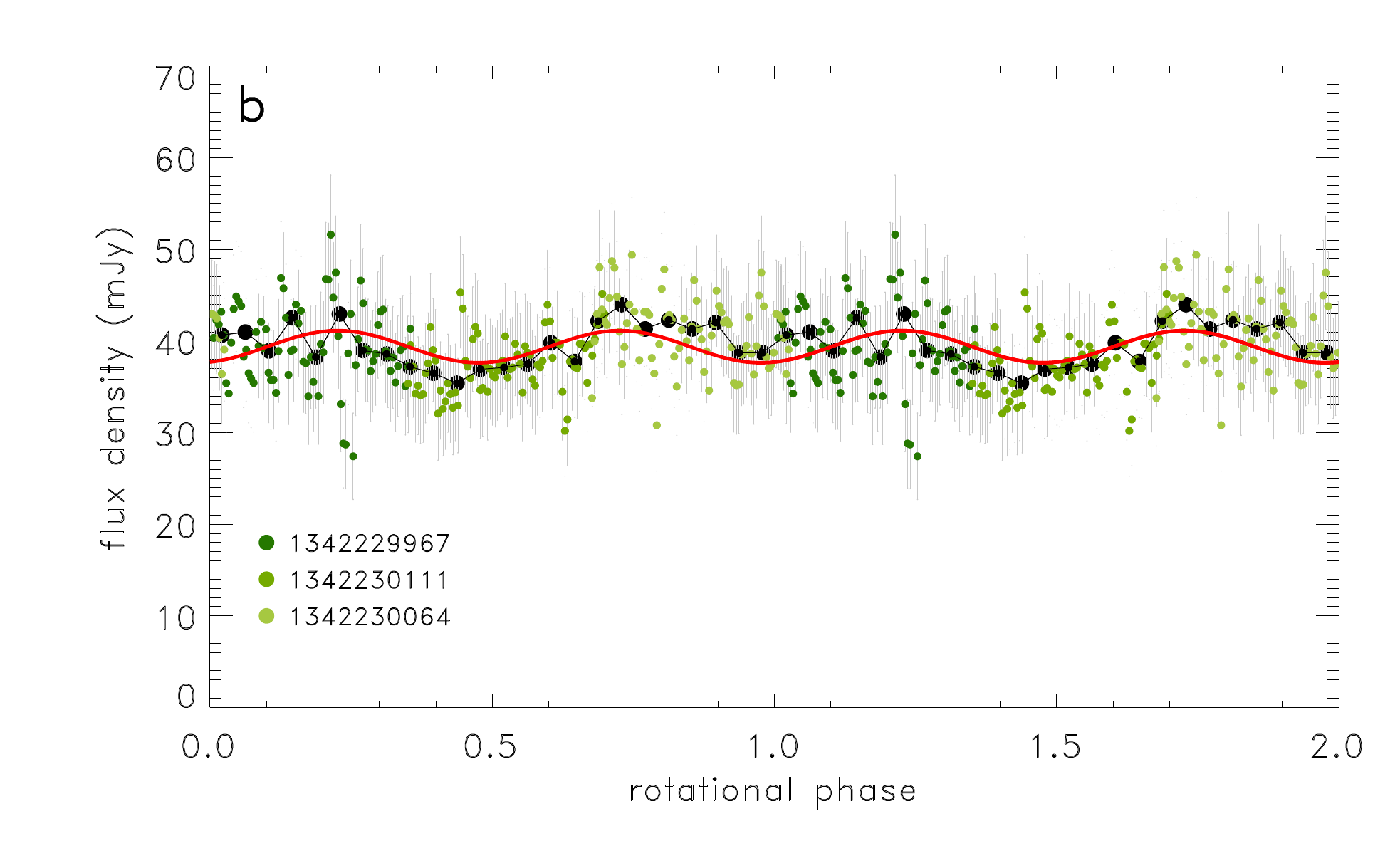}
            \includegraphics[width=0.5\textwidth]{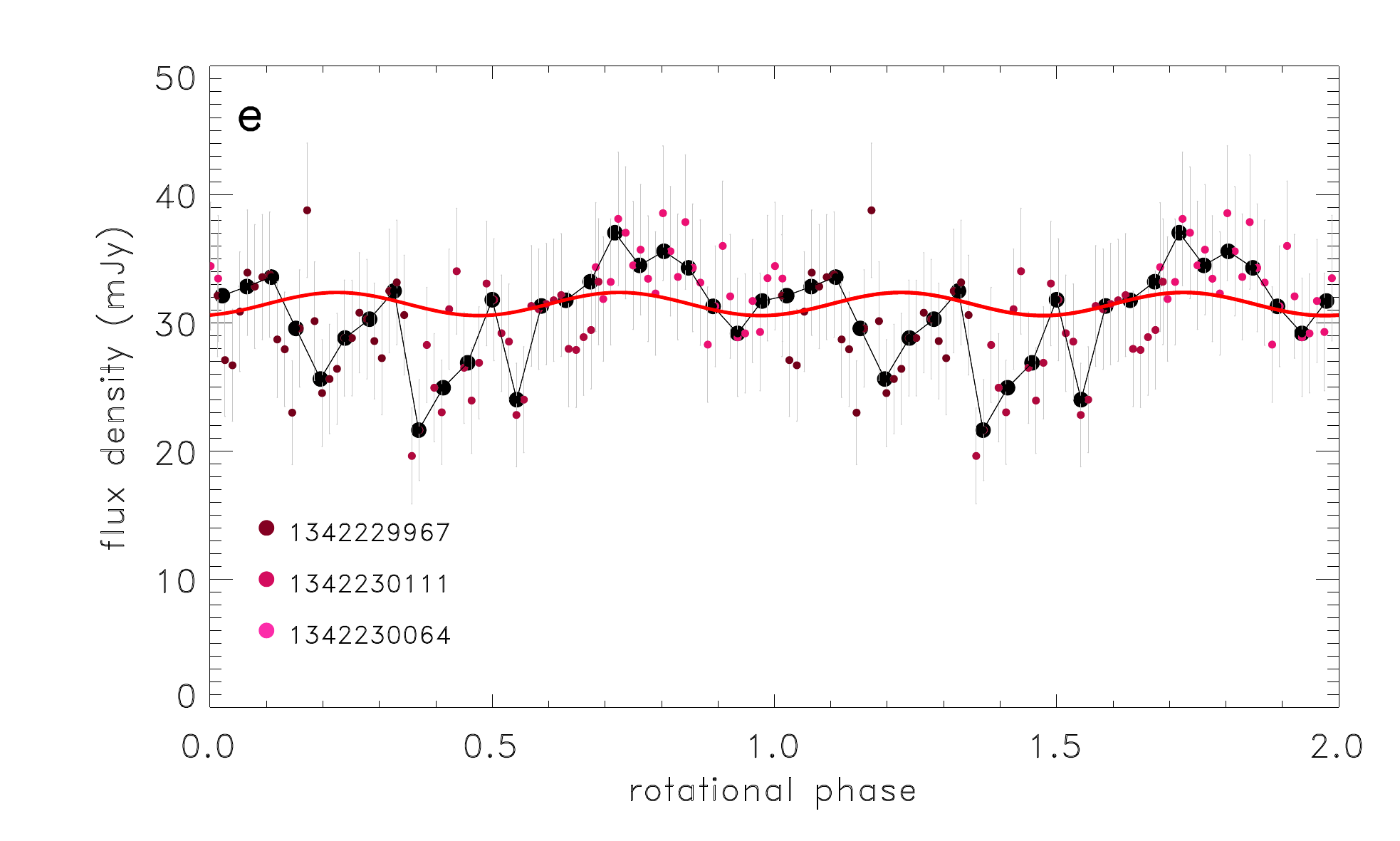} }
    \hbox{ \includegraphics[width=0.5\textwidth]{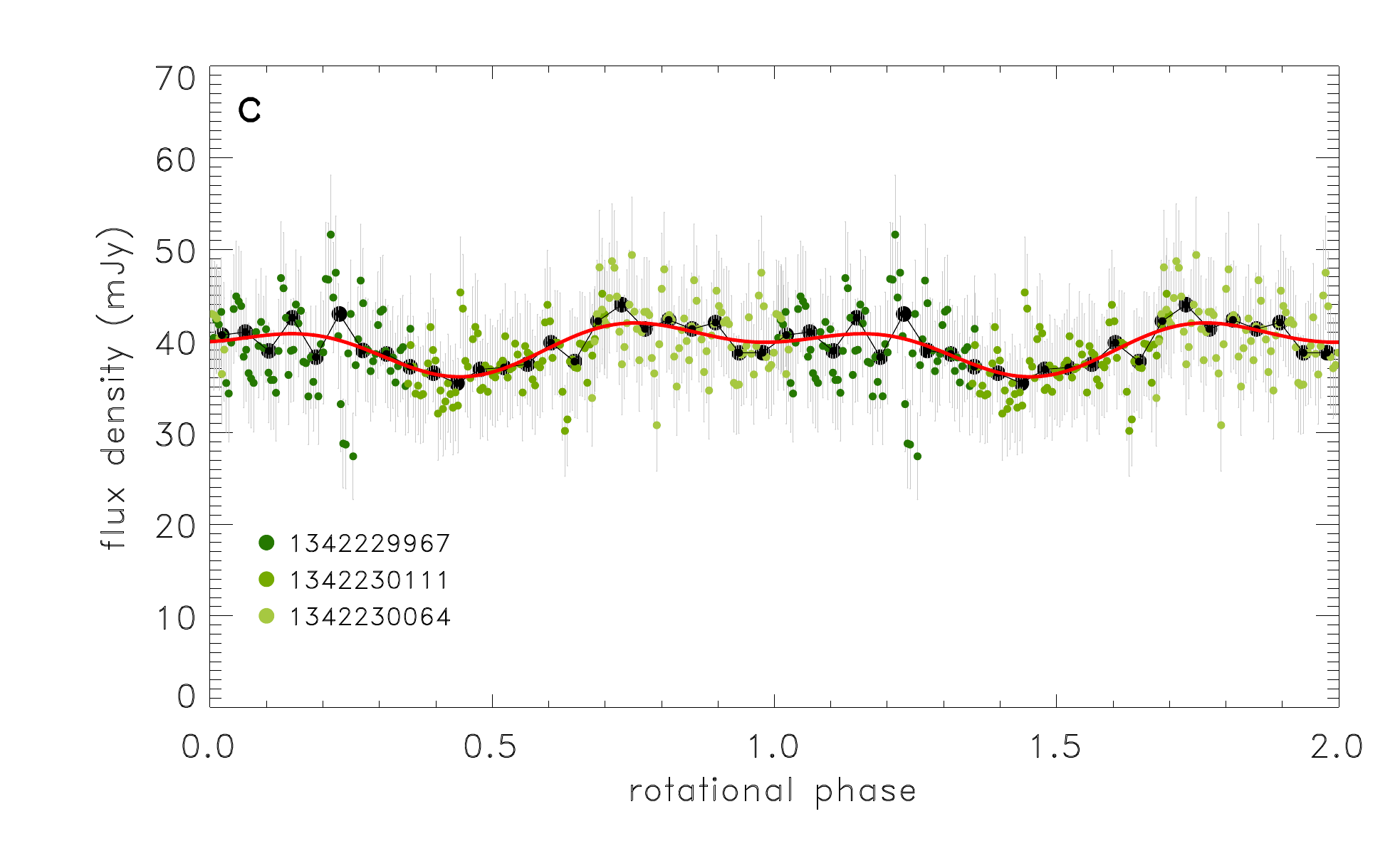}
         \includegraphics[width=0.5\textwidth]{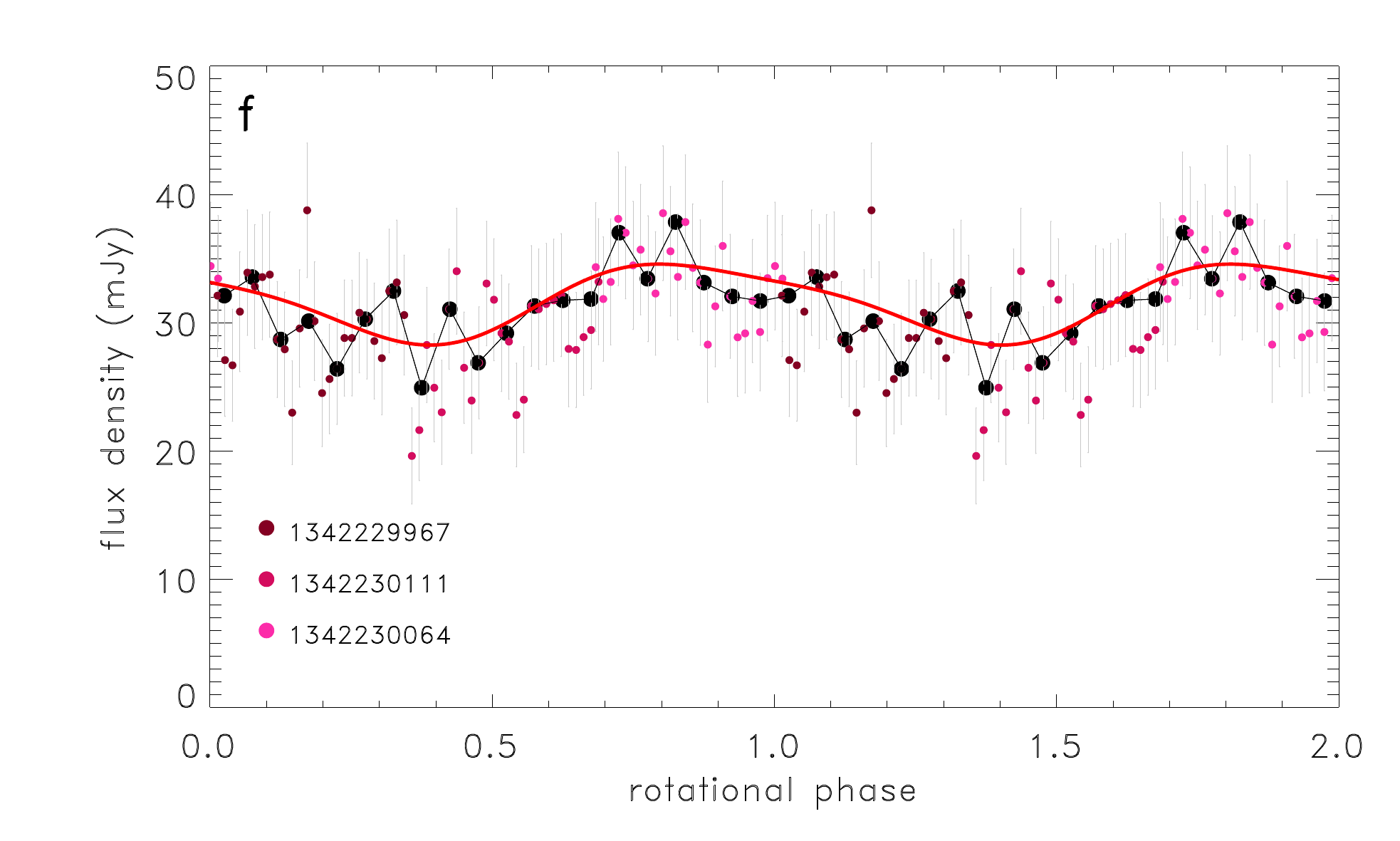}}
    \caption{Herschel/PACS 100 and 160\,$\mu$m folded light curves of Quaoar. Panel a: 100\,$\mu$m thermal light curve folded with the P\,=\,8.876\,h period. The red curve is the best-fit single-peak sine function. The black dots and curve represent the binned data. The  different coloured dots correspond to different OBSIDs, as indicated in the { legend}. Panel b: 100\,$\mu$m thermal light curve folded with the P\,=\,17.752\,h period. The red curve is a best-fit sine function, with a half-period (P\,=\,8.876\,h), and assuming equal half-amplitudes.
    Panel c: Same as b, but here the red curve is the best-fit second-order sine function, a combination of P\,=\,8.876\,h and P\,=\,17.752\,h components, i.e. allowing different amplitudes for the first and second half of the folded light curve. Panels d, e, and f:  Same as panels   a, b, and c, but using the 160$\mu$m data. }
    \label{fig:lcpacs}
\end{figure*}

Data reduction of the PACS light curve measurements were performed in the framework of the `Small Bodies: Near and Far' programme  \citep{Muller2018}, and we used the User Provided Data Products (UPDP) of these measurements available in the Herschel Science Archive.\footnote{https://archives.esac.esa.int/hsa
} The data reduction steps are described in detail in \citet{Kiss2017a}, from the reduction of raw data to the production of the background-corrected maps that { we used} for photometry. From the various UPDP data products available we used the LCDIFF maps as these have   proved to provide the best performance over both the nominal maps without background correction and the supersky-subtracted maps \citep[see][for details]{Kiss2014}. 

As in the case of other Herschel/PACS TNO  light curve targets \citep[see e.g.][]{Muller2019} a single repetition is not long enough to provide a suitable signal-to-noise ratio, and therefore the raw data of multiple repetitions are merged already in the production of the raw (background-uncorrected) maps \citep{Kiss2014,Kiss2017a}. The number of repetitions used for a single image is filter dependent; as in the case of other PACS light curve targets, we used the maps with three and six repetitions for the 100 and 160\,$\mu$m bands { (14.3 and 28.6\,min integration times, respectively)}. We used the second measurement as the first measurement's background when producing the LCDIFF files, and the second and third measurements as mutual backgrounds.  { Examples of these individual LCDIFF images are shown in Fig.~\ref{fig:lcdiff}}. The photometry was performed using the faint source optimised pipeline of the `TNOs are Cool!' Open Time Key programme { \citep{Kiss2014}}. This provided significantly more consistent flux densities than the standard Herschel Interactive Pipeline Environment \citep[HIPE,][]{Ott2010} photometry both in general in the Herschel/PACS TNO observations, and  in the specific case of the Quaoar thermal light curve measurements, both at 100 and 160\,$\mu$m \citep{Kiss2014}. 
The PACS 100 and 160\,$\mu$m photometry data are provided in Table~\ref{table:pacslcdata}. 
The colour  corrections for our PACS  light curve, and  for the standard PACS  multi-filter data are negligible as these corrections are below 2\% for objects with temperatures in the 40-50\,K range 
\citep{Muller2011}.

\begin{table*}
    \caption{Photometric data of the PACS 100 and 160$\mu$m light curve measurements}
    \begin{center}
    \begin{tabular}{ccccccccccc}
    \hline
    t$_{start}$ & t$_{end}$ & OBSID & band & SR & NM & F$_i$ & $\delta$F$_i$ & r$_h$ & $\Delta$ & $\alpha$\\
    JD & JD & & ($\mu$m) & & & (mJy) & (mJy) & (au) & (au) & (deg) \\
     \hline
2455836.29693&  2455836.30672&  1342229967&  100&001&  3&       40.35&        8.06  & 43.1141  &   43.4057  &    1.2817\\
  2455836.30020&  2455836.30999&  1342229967&  100&002&  3&       42.29&        8.25 & 43.1141  &   43.4057  &    1.2817\\
  2455836.30346&  2455836.31325&  1342229967&  100&003&  3&       41.81&        8.21 & 43.1141  &   43.4058  &    1.2817\\
   ... & ... &  ... & ... &  ... & ...   &   ...    &    ...   &  ...  &   ... &   ... \\
   2455836.29693 & 2455836.31651 &  1342229967 & 160 &  001 & 6    &   32.11    &    6.72   &  43.1141  &   43.4058  &    1.2817 \\
  2455836.30672 & 2455836.32631 & 1342229967   & 160 &  004 & 6    &   27.08    &    6.19   &  43.1141  &   43.4059  &    1.2816 \\
  2455836.31652 & 2455836.33610 & 1342229967   & 160 &  007 & 6    &   26.68    &    6.17   &  43.1141  &   43.4061  &    1.2816 \\
   ... & ... &  ... & ... &  ... & ...   &   ...    &    ...   &  ...  &   ... &   ... \\
 \hline 
    \end{tabular}
    \label{table:pacslcdata}
    \\
 \end{center}
 \footnotesize
 { Note.} { Dates are presented in spacecraft-centric reference frame. The table is available in its entirety in electronic format. The columns are: t$_{start}$, start time of block (JD); t$_{end}$, end time of the block (JD); OBSID, PACS observation ID; band, central wavelength of the photometric band; SR, starting repetiton; NM, number of repetitons merged (one repetition corresponds to 286\,s integration time); F$_i$, in-band flux density; $\delta$F$_i$, flux density uncertainty; r$_h$, heliocentric distance; $\Delta$, observer distance; $\alpha$, phase angle.}
\end{table*}

As the rotation period of Quaoar is known,  at least within the single-peaked versus double-peaked ambiguity,  we used the P$_1$\,=\,8,876\,h (single-peaked) and P$_2$\,=\,17.752\,h (double-peaked) periods, as obtained above, to fold the observed 100 and 160\,$\mu$m light curves (see Fig.~\ref{fig:lcpacs}). 
As these periods are very close to the 8.84\,h and 17.68\,h periods obtained in previous works, using one or the other period does not affect noticeably the results presented below. 

At 100\,$\mu$m the single-peaked  light curve shows a clear, closely sinusoidal variation (Fig.~\ref{fig:lcpacs}a), with a peak-to-peak amplitude of $\Delta F$\,=\,3.8$\pm$0.9\,mJy that corresponds to a $\sim$10\% flux density variation, and a maximum at $\phi$\,=\,0.46$\pm$0.05 rotational phase { ($\phi$\,=\,0 corresponds to the start of the observations)}. When using the double-peaked period, the best-fit sinusoidal with two equal half-amplitudes (Fig.~\ref{fig:lcpacs}b) has an amplitude of $\Delta F$\,=\,3.6$\pm$1.1\,mJy ($\sim$9\% flux density variation). However, in this case the folded light curve is strongly affected by a possible flux density offset between data obtained at different epochs, and that a certain rotational phase range is only covered by a single epoch. The reduced $\chi^2$ values of the single- and double-peaked cases do not differ significantly ($\chi^2_r$\,=\,0.75 and 0.76, without rescaling the uncertainties), but they are clearly different from the $\chi^2_r$\,=\,1.5 value of a constant light curve. The overall best  (lowest $\chi^2$) double-period sinusoidal fit is obtained by allowing different half-period amplitudes (Fig.~\ref{fig:lcpacs}c); however, in this case the second minimum is very shallow, which again could be caused by a small flux density offset between data from different epochs. In this sense the single-peaked folded light curve may currently provide the   best description of the 100\,$\mu$m light curve, as in this case there is a mixing of data from different epochs in rotational phases, reducing the effect of the potential offsets.  

A similar picture is obtained for the 160\,$\mu$m light curves. Here the single-peaked curve can be fitted with a sinusoidal with an amplitude of 
$\Delta F$\,=\,3.4$\pm$1.0\,mJy, or $\sim$11\% amplitude, and with a phase at maximum of $\phi$\,=\,0.53$\pm$0.05 (Fig.~\ref{fig:lcpacs}d). 
Overall, the light curve amplitude is similar to that obtained from the 100\,$\mu$m data using the same period, and also the light curve phases are similar, taking into account that there is a $\Delta\phi$\,=\,0.04 phase offset between our 100 and 160\,$\mu$m data points due to the larger number of repetitions  (six vs three) considered for 160\,$\mu$m. In the case of the double-peaked light curves (Fig.~\ref{fig:lcpacs}e and f) the flux density offsets of the epochs are quite pronounced making the equal half-period double-peaked light curve fit (Fig.~\ref{fig:lcpacs}e) very poor. While the fit is significantly better when different  half-amplitudes are allowed (Fig.~\ref{fig:lcpacs}f), this fit is still strongly dominated by the offsets between the epochs. 

\section{Additional thermal data \label{sect:addthermal}}

\begin{table*}
\caption[]{Additional Herschel/PACS and Spitzer/MIPS observations of Quaoar. } 
\small
\begin{tabular}{lllrlllrrr}
\noalign{\smallskip} \hline \noalign{\smallskip}
 &   &   & t$_{int}$  & r  & $\Delta$  & $\alpha$  & $\lambda$  & F  & $\delta F_{abs}$ \\
Instr. & ID & mid-time & [min] & [au] & [au] & [$^{\circ}$] & [$\mu$m] & [mJy] & [mJy] \\
\noalign{\smallskip} \hline \noalign{\smallskip}
PACS & 1342205970/5971/6017/6018 & 2455476.809 & 37.9 & 43.1506 & 43.56 & 1.2 &  70.0 & 32.24 & 2.11 \\
PACS & 1342205972/5973/6019/6020 & 2455476.823 & 37.9 & 43.1506 & 43.56 & 1.2 & 100.0 & 37.52 & 2.34 \\
PACS & 1342205970-5973/6017-6020 & 2455476.816 & 75.7 & 43.1506 & 43.56 & 1.2 & 160.0 & 35.62 & 2.85 \\
\noalign{\smallskip} \hline \noalign{\smallskip}
MIPS & 10676480 \& & 2453467.887 &  11.4 & 43.3455 & 42.9746 & 1.24 & 23.68 &  0.26 & 0.06 \\ 
MIPS & 10676736 & 2453467.889 &  11.4 & 43.3455 & 42.9746 & 1.24 & 71.42 & 26.99 & 4.90 \\
MIPS & 15475968/6480/7248/9552/9808 \& & 2453829.465 & 145.7 & 43.3126 & 43.1466 & 1.31 & 23.68 &  0.22 & 0.02 \\
MIPS & 15480064/0320/0576/0832/1088/1344/1600 & 2453829.470 & 145.7  & 43.3126 & 43.1465 & 1.31 & 71.42 & 28.96 & 3.67 \\
\noalign{\smallskip} \hline 
\end{tabular}
\label{table:pacsmips}                  
\\
\footnotesize
{ Note.} { The columns are: instrument, observation IDs / AOR keys; observation mid-time (Julian date); integration time; heliocentic distance, observer distance; phase angle; central wavelength of the photometric band; in-band flux density; flux density uncertainty. }
\end{table*}

In addition to the 100/160\,$\mu$m light curves, we also used one set of standard PACS multi-filter measurements. An earlier version of these flux densities was   presented in \citet{Fornasier2013}. The measurements followed the standard `TNOs are Cool!' measurement sequence \citep{Muller2009}, which means that  the target was observed at two epochs (referred to as visit-1 and visit-2), and the time between the two visits was set in a way that the target moved $\sim$30\arcsec{} with respect to the sky background that allowed us to use observations at the two epochs as mutual backgrounds. Observations at a specific visit also included scan/cross-scan observations in the same band, and in both possible PACS photometer filter combinations (70/160 and 100/160\,$\mu$m, see Table~\ref{table:pacslcdata}).
{ We note that the PACS light curve measurements used only a single scan direction and the 100/160\,$\mu$m filter combination}. Basic data reduction of the data is performed { using the same pipeline as in the case of the PACS light curve data (Sect.~\ref{sect:pacslc})}. 
The scan and cross-scan images of the PACS band are combined to produce the co-added images for each epoch. The co-added images of the two epochs are further combined to obtain background-eliminated differential (DIFF) images. The background matching method \cite{Kiss2014} is applied to correct for the small offsets in the coordinate frames of the visit-1 and visit-2 images using images of systematically shifted coordinate frames and then determining the offset that provides the smallest standard deviation of the per-pixel flux distribution in a pre-defined coverage interval;  the optimal offset can be most readily determined using the 160\,$\mu$m images due to the strong sky background with respect to the instrument noise. The differential (DIFF) images contain a positive and a negative source separated by $\sim$30\arcsec, corresponding to the two observational epochs, and these are combined in the double-differential (DDIFF) images \citep{Kiss2014} to obtain a single mean flux density at each band (see Table~\ref{table:pacsmips}). 
{ As these images combine data from multiple epochs they are an accurate representation of the mean flux densities over the observed period.}

We also used Spitzer/MIPS measurements, re-reduced via methods described in \citet{Stansberry2007,Stansberry2008} and \citet{Brucker2009}, in combination with the latest ephemeris information, and we used the flux densities provided by \citet{Mueller2012}, as shown in Table~\ref{table:pacsmips}. The Spitzer/MIPS observations are colour-corrected (divided by 1.10 and 0.89 at 23.68 and 71.42\,$\mu$m, respectively), and an estimated absolute flux calibration error of 10\% was added. 

\section{The shape and size of Quaoar}
\label{sec:occ}

\begin{figure*}
    \centering
    \includegraphics[width=0.65\textwidth]{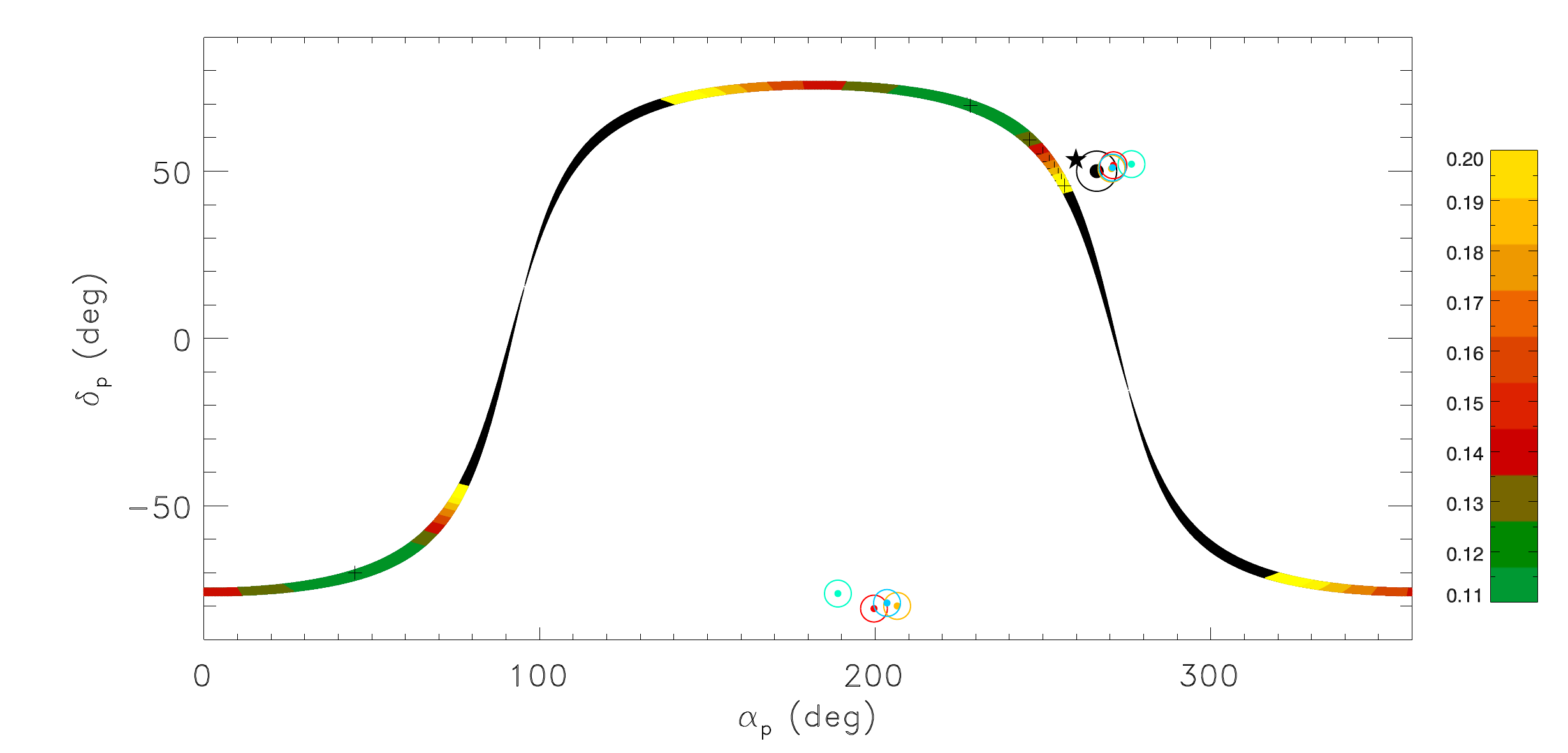}
   \caption{ Pole orientations of Quaoar's main body ($\alpha_p$,$\delta_p$), allowed by the occultation constraints, assuming an oblate spheroid shape with true oblateness values 
  { 0.11\,$\leq$\,$\epsilon$\,$\leq$\,0.2} (see the colour bar). The black regions require $\epsilon$\,$>$\,0.20. The black star is the preferred pole solution of the Q1R ring \citep{Pereira2023}. The black dot and circle mark the pole solution of Weywot's orbit by \citet{Pereira2023}. The coloured dots and circles mark the Weywot orbital pole solutions by \citet{Fraser2013} (red with highest probability). 
  { The black plus signs (+)   mark the two pole positions  used in our radiometric models: ($\alpha$,$\delta$)\,=\,(225$\degr$,+70$\degr$) and (45$\degr$,-70$\degr$).}}    
   \label{fig:oblatepoles}
\end{figure*}

The shape of Quaoar is currently best constrained by the occultation results presented in \citet{Pereira2023}. The possible occultation timing issues discussed in \citet{Braga-Ribas2013} associated with the 2011 May occultation data allow a large variety of shapes to be fitted, depending on the timing offsets considered, and therefore { provide no additional useful} constraints. { Ring detection in \citet{Morgado2023} is based on occultations in 2018-2021, most of which missed the main body of Quaoar.}
{ \citet{Pereira2023} give the position angle of Quaoar's pole ($P_Q$) and apparent oblateness ($\epsilon'$) as 354.2$\pm$1.2\,deg and 0.12$\pm$0.01, respectively, based on the 2022 August 9  occultation chords.
}
We note that the position angle of the Q1R ring is {different} from the position angle of Quaoar's main body (P$_{Q1R}$\,=\,350.2$\pm$0.2\,deg):  the two pole orientation vectors are not perfectly aligned, but they are probably rather close. We calculated the possible pole orientations of Quaoar's main body (right ascension $\alpha_p$ and declination $\delta_p$), which are compatible (within 1$\sigma$) with the observed $P_Q$ position angle { assuming that Quaoar’s shape is an oblate spheroid with a true oblateness $\geq$\,0.11 (the minimum $1\sigma$ value consistent with the occultation shape). If we further assume that the true oblateness is $\leq$\,0.20 (a reasonable but assumed upper limit), a more restricted range of on-sky pole positions are allowed, as shown by the coloured regions in Fig.~\ref{fig:oblatepoles}.}

Quaoar's apparent cross-section, based on an elliptical fit to { the chords of the 2022 occultation}, is given by a semi-major axis of a$^\prime$\,=579.5$\pm$4.0\,km and a semi-minor axis of b$^\prime$\,=\,510$\pm$9\,km ($\epsilon$\,=\,0.12$\pm$0.01), leading to an area-equivalent radius of 543$\pm$2\,km \citep{Pereira2023}. However, they also found a 12\,km difference to an earlier result by \citet{Braga-Ribas2013}, which could indicate a triaxial ellipsoidal shape instead of the oblate spheroid { assumed above}. { The triaxial shape is also supported by the preferred double-peaked light curve (see Sect.~\ref{sect:k2}). Any triaxial shape and the related pole solution have to match the apparent shape at the time of the occultation, as given above. } 
For our radiometric study we consider both oblate spheroid and triaxial ellipsoid shape solutions in Sect.~\ref{sec:tpm}.

\section{Thermal emission estimates of Quaoar's rings \label{sect:ring}}
\label{sec:rings}

We used the simple ring model previously applied by \citet{Lellouch2017} and \citet{Muller2019} to analyse the thermal emission of the Chariklo and Haumea rings. A detailed description of the model and the related references can be found in \citet{Lellouch2017}, we just provide the main equations here. 
In this model the ring is assumed to be infinitely thin, and characterised by the radius, width, visible range normal optical depth ($\tau$), and reflectivity (I/F) of the grains. { Reflectivity is related to the Bond albedo $A_r$ via a the phase integral $q$ ($A_r$\,=\,$q\times$I/F), which is assumed to be q\,=\,0.5 \citep[see][]{Muller2019}.} The temperature of the ring particles is obtained as
\begin{equation}
    T_p = \Bigg( \frac{(1-A_r) \sin{B_\odot} (1-e^{-\tau/\sin B_\odot}) S_\odot}
    {(1-e^{-\tau})\epsilon_{r,b} \, \sigma \, f \big(1 - \frac{6(1-e^{-\tau})}{4\pi} \big) r_h^2}  \Bigg)^{1/4}
,\end{equation}
where $A_r$ is the Bond albedo, $B_\odot$ is the solar elevation above the ring plane,  $\tau$ is the optical depth, $S_\odot$ the solar constant, $\epsilon_{r,b}$ is the bolometric emissivity of the ring particles, $\sigma$ is the Stefan-Boltzmann constant, $f$ is the rotation rate factor (assumed to be f\,=\,2 here), and $r_h$ is the heliocentric distance. The brightness temperatures are obtained as
\begin{equation}
    B_\nu (T_B(\lambda)) = \epsilon_{r,\lambda} (1-e^{-\tau/\sin B})B_\nu(T_p)
,\end{equation}
where $B_\nu (T)$ is the Planck function, $\epsilon_{r,\lambda}$ is the spectral emissivity, and  $B$ is the elevation of the observer above the ring plane, and it is assumed that  $\epsilon_{r,\lambda}$\,=\,$\epsilon_{r,b}$\,=\,1.

\begin{figure}[ht!]
    \centering
    \includegraphics[width=\columnwidth]{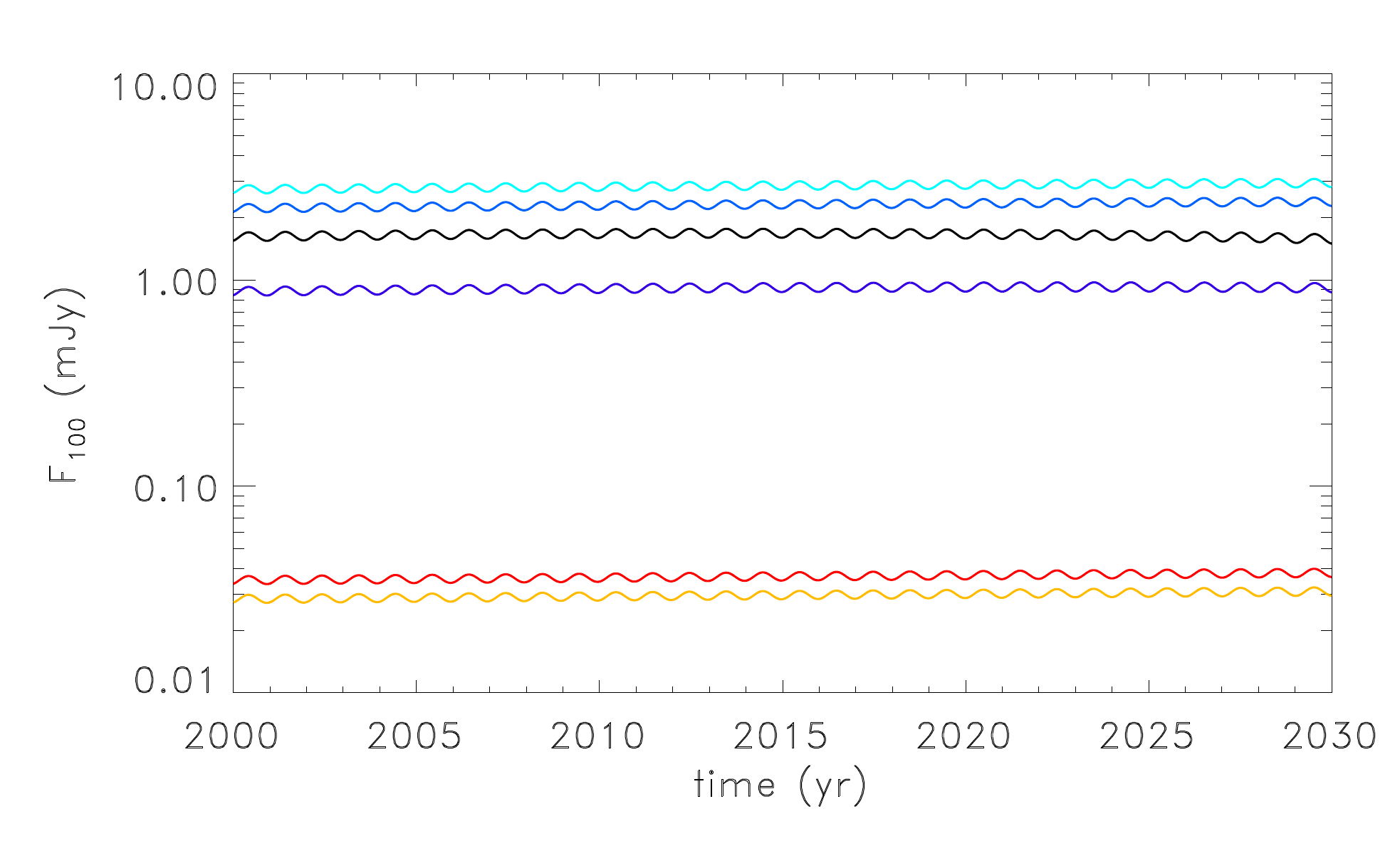}
    \caption{Flux density estimates of Quaoar's rings for the period 2000-2030 at 100\,$\mu$m. The curves correspond to homogeneous ring models with the following parameters (ring ID, reflectivity, optical depth, width, and radius): 
    light blue, [Q1R, I/F\,=\,0.04, $\tau$\,=\,0.04, w\,=\,52\,km, r\,=\,4057\,km]; 
     blue, [Q1R, 0.50, 0.04, 52, 4057]; 
    black: [Q1R, 0.04, 0.40, 5.2, 4057];
    purple: [Q1R, 0.50, 0.40, 5.2, 4057];
    red: [Q2R, 0.04, 0.004, 10, 2520]; 
    orange: [Q2R, 0.50, 0.004, 10, 2520]. }
    \label{fig:ringfd}
\end{figure}
To estimate the thermal emission from Quaoar's rings we took the ring parameters listed in Table~E.1 in \citet{Pereira2023}. 
An upper limit for the possible thermal emission contribution can be obtained assuming that the denser Q1R ring is predominantly made of material similar to that in its `dense' sections, which are characterised either by a wider ring with intermediate optical depth (w\,=\,52\,km, $\tau$\,=\,0.04) or by a narrow ring with a high optical depth (w\,=\,5.2\,km, $\tau$\,=\,0.4). The flux density estimates show (Fig.~\ref{fig:ringfd}) that the wider ring with the intermediate optical depth would give the higher flux density from these two scenarios. Depending on the reflectivity chosen these may result in flux densities of 1--3\,mJy at 100\,$\mu$m (i.e. $\sim$3-8\% of the total thermal emission of the system).  
We also calculated an  average  Q1R ring thermal emission estimate considering all ingress and egress Q1R ring detections listed in Table~E.1 in \citet{Pereira2023}, which covered ring sections with different widths and optical depths. We assumed that this sample is representative of the frequency of ring sections with different properties, and calculated the mean flux densities for I/F\,=\,0.04 and 0.5.  We obtained 1.4 and 1.13 mJy at 100\,$\mu$m (i.e. $\sim$3\% of the total flux density of the system for a wide range of I/F reflectivity values). 
As the weak Q2R ring is much more homogeneous \citep{Pereira2023}, it is more straightforward to estimate its thermal emission contribution. Assuming r\,=\,2520\,km, w\,=\,10\,km, $\tau$\,=\,0.004, and I/F\,=\,0.04 and 0.5 (red and orange curves in Fig.~\ref{fig:ringfd}), we obtain $\sim$0.03\,mJy at 100\,$\mu$m (i.e. $\sim$0.1\% of the total thermal emission of the system). Estimates for the thermal emission contribution of the dominant Q1R ring at different wavelengths are given in Table~\ref{table:ringfd}. 
\begin{table}[ht!]
        \caption{Monochromatic flux density estimates of Quaoar's Q1R ring and Weywot.}
    \small
    \begin{tabular}{c|ccc|ll}
    \hline
    $\lambda$ & Q1R$_{max}$ & Q1R$_{avg}$ &  Q1R$_{avg}$ & W$_{mean}$ & W$_{max}$\\
              &             & I/F\,=\,0.04 & I/F\,=\,0.5 &   &    \\
    ($\mu$m) & (mJy) & (mJy) & (mJy) & (mJy) & (mJy) \\
    \hline
    24  &   0.0217 &     0.0095  &    0.0042 & 0.012 & 0.024 \\
    70  &   2.3338 &     1.1217  &    0.8426 & 0.68  & 1.32 \\
    100 &   2.8607 &     1.3962  &    1.1342 & 0.78  & 1.50 \\
    160 &   2.3220 &     1.1473  &    0.9918 & 0.60  & 1.16 \\
    250 &  1.4189  &    0.7057   &   0.6304  & 0.36  & 0.70 \\
    350 &  0.8761  &    0.4371   &   0.3964  & 0.22  & 0.43 \\
    500 &  0.4925  &    0.2463   &   0.2257  & 0.12  & 0.24 \\
    870 &  0.1855  &    0.0930   &   0.0860  & 0.046 & 0.089 \\
   1300 &  0.0880  &    0.0441   &   0.0410  & 0.022 & 0.042 \\
   \hline
    \end{tabular}
    \label{table:ringfd}
    \footnotesize
    \\
    { Note.} { Flux density estimates are presented here for some relevant infrared wavelengths, including those of Spitzer/MIPS (24 and 70\,$\mu$m), Herschel/PACS (70, 100, and 160\,$\mu$m),   SPIRE (250, 350, and  500\,$\mu$m), and ALMA (bands 7 and 6, 870 and 1300\,$\mu$m). Q1R$_{max}$ is the expected maximum contribution assuming a Q1R ring with w\,=\,52\,km, $\tau$\,=\,0.04, I/F\,=\,0.04 (main text). The  Q1R$_{avg}$ columns give the expected average contributions, assuming I/F\,=\,0.04 and I/F\,=\,0.5. The Weywot flux estimates (W$_{mean}$ and W$_{max}$) are obtained via NEATM, assuming a beaming parameter $\eta$\,=\,1.2. }
\end{table}

While the contribution of Q2R is certainly negligible, Q1R is expected to have a small but notable contribution. 
In extreme cases, when most of the Q1R is dense, the total contribution of the Q1R may be as high as $\sim$8\% of the total thermal emission. A more sophisticated radiative transfer model and a much better mapping of the Q1R ring structure are needed to give a more accurate estimate for the ring's contribution to the thermal emission of the system. 

\section{Thermal emission estimate for Weywot}
\label{sec:weywot}

Quaoar's satellite Weywot was discovered in 2007 \citep{BrownSuer2007}. It is 5.6$\pm$0.2\,mag fainter than its primary \citep{Fraser2010,BB17}. With the assumption of an equal albedo, this would give a Weywot/Quaoar size ratio of 1/12 or about 70-100\,km diameter for the satellite \citep[e.g.][]{Fornasier2013}. From the occultation results \citep{Braga-Ribas2013} it was concluded that Weywot is orbiting outside the equatorial plane of Quaoar. 
It is worth   noting here that the radiometric results based on existing infrared data \citep[e.g. in][]{Fornasier2013}, as well as quantities derived from light curve and absolute photometric measurements in reflected sunlight, refer to Quaoar-Weywot system properties, while stellar occultations can usually only { reveal} a single body. For our study it is important to consider both objects individually.
From a radiometric point of view, very little is known about Weywot. The estimates from the 5.6\,mag brightness difference give a size below $\sim$100\,km, while occultations (single chord from 04 August 2019 reported by \citet{Kretlow2019}; multi-chord event measured in June 2023, Fernandez-Valenzuela, in prep.) point to a much larger size, between 170 and 200\,km. 
{ Using a simple NEATM radiometric model approach \citep[see e.g.][]{Harris1998,Lellouch2013} a 100\,km Weywot with  geometric albedo of p$_V$=0.1, beaming parameter of $\eta$\,=\,1.2, at a heliocentric distance of r\,=\,43.3\,au, and an observer distance of $\Delta$\,=\,43.0\,au} would produce roughly 0.007, 0.40, 0.46, and 0.36\,mJy at 24, 70, 100, and 160\,$\mu$m, respectively. This would be negligible as Quaoar is about 50-100 times brighter in the infrared. A 200 km body at the same distance, but with a much lower geometric albedo (p$_V$\,=\,0.05) would, however, already give 0.03, 1.63, 1.86, and 1.44\,mJy, corresponding to 12, 6, 5, and 5\% of the measured flux densities. 
In Table~\ref{table:ringfd} we present Weywot's flux densities at the relevant wavelengths for an intermediate case with D\,=\,130\,km and a geometric albedo p$_V$\,=\,0.08, and for a large and dark case with D\,=\,180\,km and p$_V$\,=\,0.05. Both settings are compatible with observed brightness values and early occultation estimates. { Here we used a beaming parameter of $\eta$\,=\,1.2 in all cases, a representative value among TNOs, as obtained by \citet{Stansberry2008}, very similar to that obtained by   \citet{Lellouch2013}}. 

\section{Thermophysical model study of Quaoar}
\label{sec:tpm}

\begin{figure}[ht!]
    \centering
    \includegraphics[width=\columnwidth]{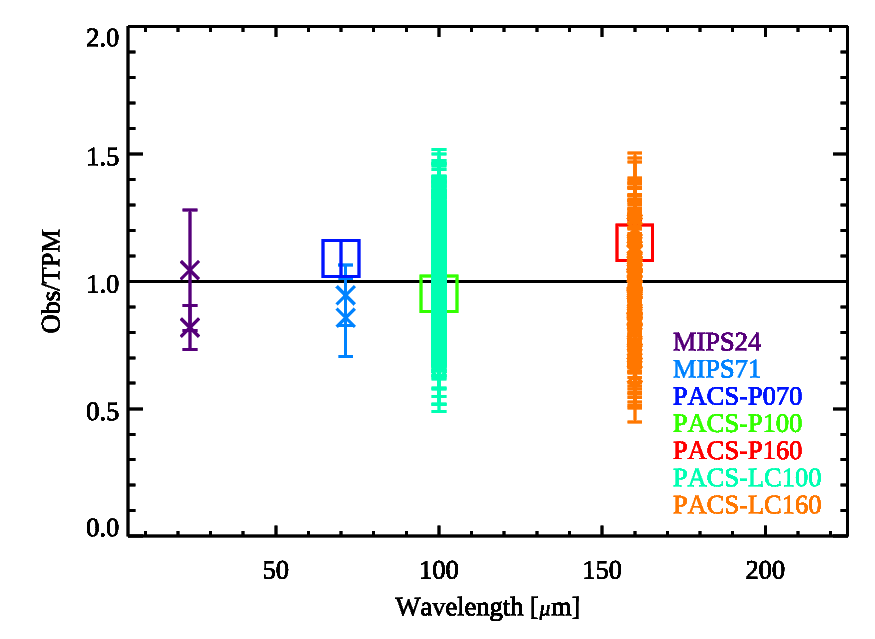}
    \includegraphics[width=\columnwidth]{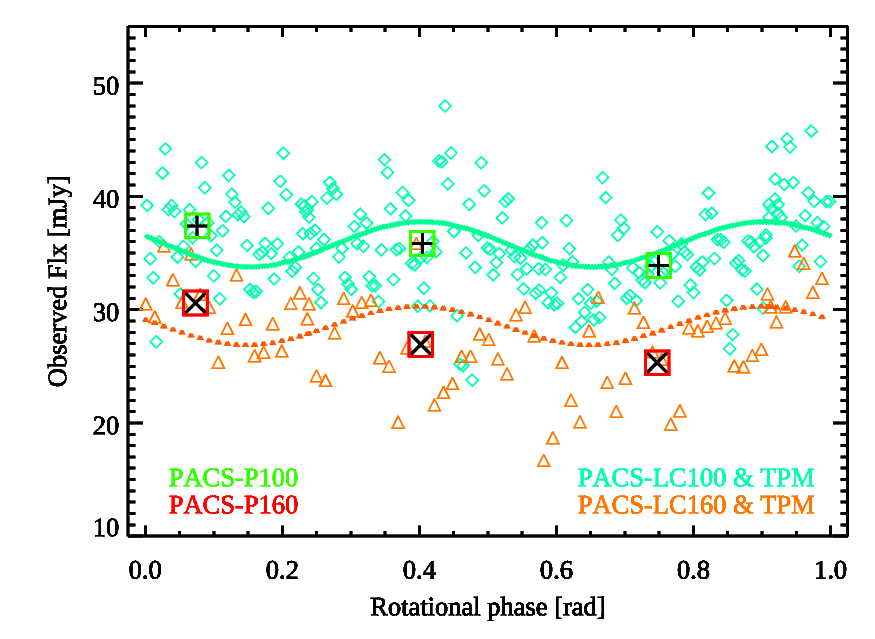}
    \caption{Thermophysical model results for Quaoar. Top: All available IR observations divided by the corresponding TPM predictions, using
    a triaxial shape with an equivalent diameter of 1090 km, and a thermal inertia of 12\,\tiu.
    The calculations take small flux contributions from the ring and Weywot into account (intermediate
    case). These absolute ratios are very sensitive to the size and thermal inertia assumptions in the model;
    the 24\,$\mu$m point is strongly influenced by surface roughness effects.
    Bottom: Small ring and Weywot flux contributions are subtracted from the observed PACS 100 
    and 160\,$\mu$m light curve data. The triaxial Quaoar model solutions (at 100\,$\mu$m in green
    and at 160\,$\mu$m in orange) { are superimposed}. The boxes represent the mean PACS values derived from
    the three observing blocks (see Table~\ref{table:lc1}). }
    \label{fig:pacs_tpm}
\end{figure}

The { thermophysical model} (TPM) by \citet{Lagerros1996A&A...310.1011L,Lagerros1997A&A...325.1226L,Lagerros1998A&A...332.1123L} and M\"uller \& Lagerros (1998, 2002) was used to interpret the available thermal measurements (see Tables~\ref{table:lc1} and \ref{table:pacslcdata} and Fig.~\ref{fig:pacs_tpm}), with possible ring and satellite contributions subtracted (Table~\ref{table:ringfd}). 
We did not consider the SPIRE \citep{Fornasier2013} and ALMA measurements \citep{BB17} as the influence of Quaoar's very uncertain effective emissivity at millimetre-submillimetre wavelengths makes it difficult to use these data directly for constraining size, shape, or thermal properties.
Our model takes the object's true illumination geometry (r, $\Delta$) and the telescope-centric observing geometry (phase angle, aspect angle, rotational phase) into account. Spin-shape solutions from simple prime-axis rotating spheres to highly complex irregular, tumbling, or multiple bodies can be calculated. { Thermally relevant surface properties are thermal inertia and surface roughness, both of which influence the object's thermal emission spectrum.}
{ The visual wavelength Bond albedo is the most influential parameter in the TPM because it determines the fraction of sunlight absorbed. To calculate it for Quaoar we started with the absolute visual magnitude from \citet{Rabinowitz2007}, $H_V$\,=\,2.729$\pm$0.025.}
{ We then subtracted the most likely visible-light contribution from the rings (5\%; see Sect.~\ref{sect:ring}) and Weywot \citep[5.6$\pm$0.2 mag fainter than Quaoar;][]{Fraser2010,Vachier2012}, obtaining for the absolute magnitude of Quaoar itself H$_V$\,=\,2.79$\pm$0.35.}

NASA's New Horizons LORRI instrument observed several large KBOs at extreme phase angles that are not accessible from the ground \citep{Verbiscer2022}. For Quaoar, these measurements, combined with the low-phase angle data \citep{Rabinowitz2007}, { were used to establish the phase curve from 0.51$^{\circ}$ to 94$^{\circ}$.}
This phase curve was then used to estimate the object's phase integral (q$_V$ = 0.52 $\pm$ 0.01).
{ Quaoar's area-equivalent radius, 543$\pm$2\,km, leads to}
a geometric V-band albedo of p$_V$\,=\,0.11 ($\pm$0.01) and a Bond albedo of A$_B$\,=\,p$_V\cdot$q$_V$\,=\,0.057$\pm$0.006.

The TPM allows us to test different scenarios for Quaoar itself, but also for the ring system and Weywot. 
For the rings (see Sect.~\ref{sec:rings}) and Weywot (see Sect.~\ref{sec:weywot}) we { considered} three explicit cases: 
\begin{itemize}
    \item { RW1}: Weywot and the rings do not contribute significantly to the system-integrated infrared fluxes; 
    \item { RW2}: Weywot has an intermediate size (130 km diameter) and albedo (p$_V$=0.08), and we assume the average Q1R contribution (Q1R$_{avg}$, I/F=0.04); 
    \item { RW3}: Weywot has a very large size (180 km diameter), combined with a dark albedo (p$_{V}$=0.05), and we assumed the maximum Q1R contribution (Q1R$_{max}$).
\end{itemize}

For the primary body we assumed oblate spheroids and triaxial bodies of different
equivalent sizes (size of an equal-volume sphere), but within the given constraints from the occultation
results: 

\begin{itemize}

    \item { OB1:} Oblate spheroid.  { We consider the occultation result by \citep{Pereira2023} 
    as a limiting case with the smallest possible oblateness. This corresponds to an oblate spheroid with
    a\,=\,b\,=\,579.5\,km, c\,=\,510\,km, and $\epsilon$\,=\,0.12, and } with the Q1R spin-pole solution
    ($\lambda$, $\beta$)$_{ecl}$ = (244.25$^{\circ}$, 75.98$^{\circ}$). The equivalent diameter is $D_{eq}$\,=\,1110.65\,km and 
    the albedo is p$_V$=0.11 (connected to H$_V$=2.79 mag).
    \item { OB2:} { Oblate spheroid with $\epsilon$\,=\,0.16, the oblateness value corresponding the allowed pole solution closest to the Q1R spin-pole solution (see Fig.~\ref{fig:oblatepoles}), a\,=\,b\,=\,579.5\,km, and c\,=\,486.8\,km (D$_{eq}$\,=\,1093.6\,km, p$_V$\,=\,0.11), and using the Q1R pole. }
        \item { TA}: Triaxial shape, the 2022 occultation event happened during light curve maximum.
        An oblateness of $\epsilon$\,=\,0.16 brings Quaoar's pole closest to Q1R's preferred
pole position (see Fig.~\ref{fig:oblatepoles}). In addition we used an a/b ratio of 1.16, which is needed to match the observed visible and thermal light curve amplitudes. The equivalent size is D$_{eq}$\,=\,1040.9\,km (p$_V$=0.125).
        \item { TB}: Triaxial shape, the 2022 occultation event happened during light curve minimum.
        The same settings as listed for {\sl TA}, but now with an equivalent size of 1149.2 km (p$_V$=0.102).
        \item { TC}: Triaxial shape as in {\sl TA} and {\sl TB}, but with variable equivalent size.
    
\end{itemize}

{ The oblate spheroid (OB1 and OB2) cases that allow for a volume-equivalent size of 1110$\pm$10 km ($\epsilon$\,=\,0.12) or 1094$\pm$10\,km ($\epsilon$\,=\,0.16) work reasonably well and can explain the observed absolute flux densities from Spitzer and Herschel. The derived thermal inertias (TIs) range from 5 to $\sim$30\,\tiu. A faster rotating Quaoar (P\,=\,8.876\,h instead of 17.752\,h) would require 10-20\% lower TIs in
the TPM calculations to match the observed fluxes. The lower TIs  $\lesssim$10\,\tiu\, are found in case { RW1} where Weywot and the rings do not contribute to the system’s observed fluxes. The highest TIs are connected to case { RW3} with maximum satellite
and ring contributions. The OB1 case ($\epsilon$\,=\,0.12), with a volume-equivalent size of 1110 km, requires Quaoar's TI to be close to 30\,\tiu, while   the OB2 case ($\epsilon$\,=\,0.16, smaller size of 1094\,km), in combination with a TI of about 25\,\tiu,   matches the observations at an acceptable level. These TIs are in good agreement with the published TIs for other large TNOs or Saturnian satellites 
\citep[see Table 2 in][]{Muller2020}. However, the highest TI range above 20\,\tiu\ seems to be too large for Quaoar being located beyond 40\,au from the Sun. As a consequence, the extreme ring--Weywot case (RW3) can be excluded with high probability.}

The TPM flux predictions for the Spitzer/MIPS 24\,$\mu$m band are very sensitive to the object's surface roughness. We find that a low surface roughness with r.m.s.\ values $\le$\,0.2 are needed to bring the observations and model into agreement. The effective sizes of the oblate spheroids reproduce the observed absolute flux levels very well, indicating that this value must be close to Quaoar's true value. However, a single-albedo oblate spheroid cannot explain the visible and thermal light curve amplitudes (an oblate body at a given aspect angle does not produce a changing cross-section as it rotates). A triaxial body is needed to explain all available observations.

\smallskip

The `small' triaxial solution ({ TA}), assuming that the 2022 occultation happened during the light curve maximum, has clear difficulties  in reaching the measured infrared flux densities. A match can only be found for dark rings and for a large and dark Weywot ({ RW3}). Even then, Quaoar would be required to have a very low TI below 5\,\tiu, combined with a very smooth surface. This makes the ({ TA})  small  triaxial solution unlikely.

\smallskip

The `large' triaxial solution ({ TB}), assuming that the 2022 occultation happened during the light curve minimum, is compatible with the IR data only when the rings and Weywot have   very  little or no flux contribution ({ RW1}). In these cases, Quaoar's TI is found to be $\ge$20\,\tiu\, which is higher than TI values found for comparable objects. The comparison between infrared data and TPM predictions indicates that the assumed effective diameter
for Quaoar of 1149.2\,km is too large.

\smallskip

A triaxial shape ({ TC}), with the Q1R spin-pole, and having a mean equivalent size of 1090\,km produces the most convincing solution (with reduced $\chi^2$ values close to 1.0). The TPM predictions agree, on an absolute level, very well with all observed infrared flux densities. At the same time, the triaxial shape reproduces the visual and thermal light curves, and it is compatible with the published occultation constraints. The pure radiometric size determination for such a triaxial shape has a formal uncertainty of about $\pm$40\,km, connected mainly to the unknown flux contribution of the ring(s) and Weywot. A negligible ring and satellite flux ({ RW1}) would push the radiometric size to values above about 1100\,km, in combination with a thermal inertia below about 10\,\tiu. In the case of the expected maximum ring contribution in combination with a large and dark Weywot ({ RW3}), the radiometric size would shrink below about 1080\,km and the  thermal inertia would go up to values above about 10\,\tiu. Both triaxial radiometric solutions would fit the PACS light curves (also the visual light curves) and, due to the unknown rotational phase, both fitted occultation ellipses presented by \citet{Braga-Ribas2013} and \citet{Pereira2023}.

{ In summary,} our calculations point to a triaxial shape with a volume-equivalent size of 1090$\pm$40\,km as the most likely solution (case { TC}). In this case our TPM set-up requires a thermal inertia of 12\,\tiu\ in combination with  low surface roughness to explain the full set of thermal measurements, which is in good agreement with the properties found by \citet{Fornasier2013} (see also Fig.~\ref{fig:pacs_tpm}).

\section{Discussion and conclusions}

In this paper we presented the Kepler-K2 visible range and Herschel/PACS thermal emission light curves of (50000) Quaoar. Comparing the single- and double-peaked K2 light curves (periods P\,=\,8.876\,h and 17.752\,h), we show { in Sect.~\ref{sect:k2}} that the double-peaked light curve is { strongly} preferred, but a single-peaked light curve cannot be fully excluded. { If Quaoar is an oblate spheroid the double-peaked case would require two similar albedo features approximately 180$\degr$ apart in longitude. }
A thermal emission light curve is detected with an amplitude of $\sim$10\%   in the 100 and 160\,$\mu$m Herschel/PACS bands assuming these rotation periods. However, { the thermal light curve is more easily explained by the rotation of a triaxial ellipsoid. Such a shape is also consistent with the fact that Quaoar's figure obtained from the 2011 and 2022 occultations were detectably different \citep{Braga-Ribas2013,Pereira2023}, essentially ruling out an oblate (non-triaxial) shape.}

It is generally accepted that the largest Kuiper belt objects are in hydrostatic equilibrium, and therefore should be nearly spherical  or flattened to a shape of a Maclaurin spheroid in the simplest case assuming strengthless fluid due to slow or moderately fast rotations. 
While fast rotation is the likely cause of Haumea's triaxial shape \citep[P\,=\,3.9\,h][]{Ortiz2017}, the rotation of Quaoar (either with a period of 8.88 or 17.7\,h) is too slow to be responsible for such a distorted shape. Among the largest Kuiper belt objects Pluto and Charon were found to be very round \citep{Nimmo2017} with very small asphericity.  
Recent works using high-resolution imaging \citep{Vernazza2021}
show that main-belt asteroids with D\,$\geq$\,400\,km are spherical, with deviations within 1\% of their radii, and the transitional size from  irregular  to  spherical  objects may be as low as D\,$\approx$\,300\,km. This transitional size is expected to be even lower for the icy objects in the trans-Neptunian region due to their lower compressive strength \citep{PotatoRadius}. However, a recent study by \citet{K23} found that TNOs showed notably higher light curve amplitudes at large sizes (D\,$\geq$\,300\,km) than found among main belt asteroids. { We suggest that Quaoar may have originally been rotating fast enough to have obtained a triaxial shape (similar to Haumea), and that the shape was `frozen in'. Tidal interactions with Weywot could then have slowed the rotation to its current value, although detailed modelling of that scenario is outside the scope of this paper.}

Density estimates of Quaoar have evolved significantly due to the improvements is the determination of the satellite orbit and hence system mass, and also due to the advances in shape and size determination. \citet{Fraser2010} obtained a system mass of M$_{sys}$\,=\,1.6$\times$10$^{21}$\,kg, and an extremely high bulk density of $\rho$\,=\,4.2\,\gcc, due to the small size estimate obtained as the average of the  Spitzer \citep{Stansberry2008,Brucker2009} and Hubble \citep{Brown2004} size estimates. \citet{Vachier2012} obtained a similar system mass of 1.65$\times$10$^{21}$\,kg. Additional satellite orbit data led to system mass values of $\sim$1.3-1.5$\times$10$^{21}$\,kg \citep{Fraser2013}, with more likely solutions having the lower mass estimates of ([1.28-1.30]$\pm$0.09)$\times$10$^{21}$\,kg. 
{ Using Herschel/PACS thermal infrared measurements, Fornasier et al. (2013) obtained an updated size of D$_{eq}$\,=\,1074$\pm$38\,km. Combining this value with a system mass of 1.4$\times$10$^{21}$\,kg, they calculated the system density as $\rho$\,=\,2.13$\pm$0.29\,\gcc.}
{ Applying the infrared flux densities by \citet{Fornasier2013}, and} using additional ALMA band-6 and band-7 measurements, \citet{BB17} found D$_{eq}$\,=\,1079$\pm$50\,km, and a density estimate of $\rho$\,=\,2.13$\pm$0.29\,\gcc\, { with a system mass of 1.4$\times$10$^{21}$\,kg}. 
The latest system mass value of (1.2$\pm$0.05)$\times$10$^{21}$\,kg was obtained by \citet{Morgado2023}. \citet{Pereira2023} obtained an area-equivalent diameter of 1086$\pm$4\,km, and calculated a bulk density of 1.99\,\gcc. This is the same as obtained by \citet{Braga-Ribas2013} from another occultation event, using a mean value of M$_{sys}$\,=\,1.4$\times$10$^{21}$\,kg from \citet{Fraser2013} for the system mass, { and an equivalent diameter of 1110$\pm$5\,km obtained from the best-fit occultation chords assuming a Maclaurin spheroid shape. }

\begin{figure}[ht!]
    \centering. 
    \includegraphics[width=\columnwidth]{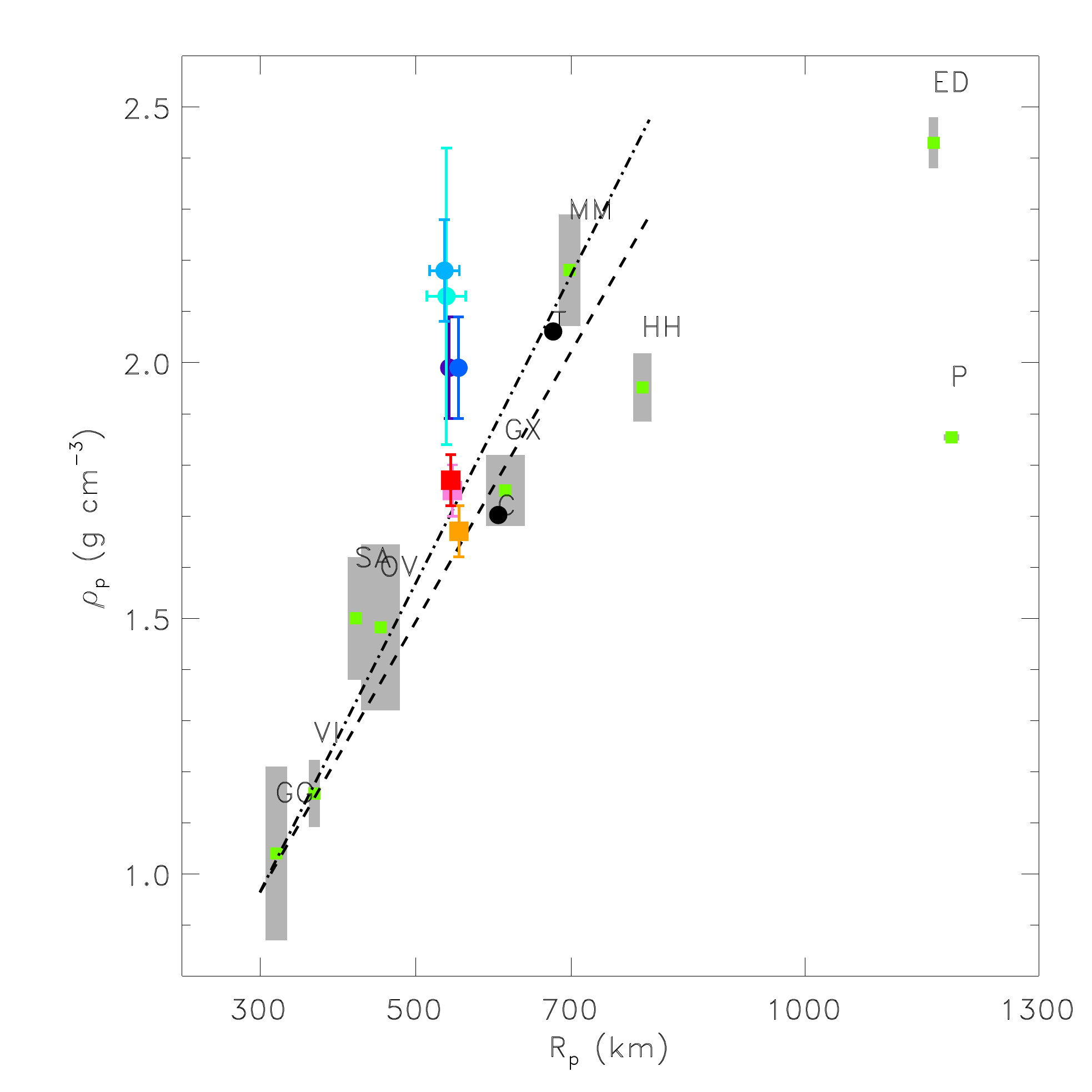}
    \caption{Bulk density vs   size (volume equivalent radius) of the primary in the ten most massive binaries in the Kuiper belt. The binaries are labelled  using the primary-secondary initials: P and C, Pluto and Charon \citep{Nimmo2017}; ED, Eris-Dysnomia \citep{Sicardy2011,Holler2021Eris}; HH, Haumea-Hi'iaka \citep{Ortiz2017,Dunham2019}; MM, Makemake-MK2 \citep{Ortiz2012,Brown2013,Parker2018}; GX, Gonggong-Xiangliu \citep{Kiss2019}; OV, Orcus-Vanth \citep{BB18}; SA, Salacia-Actea \citep{BB17}; VI, Varda-Ilmar\"e \citep{Grundy2015,Souami2020}; GG, \gg-\ggg\, \citep{Grundy2019ggg}.
    The grey boxes represent the uncertainties. 
    The different size and density estimates for Quaoar are  colour-coded, as follows: cyan: \citet{BB17}; light blue: \citet{Fornasier2013}; dark blue: \citet{Braga-Ribas2013}; purple: \citet{Pereira2023};  orange: this work, TPM model { OB1}; { pink: this work, TPM model OB2}; red: this work, TPM model { TC} (triaxial ellipsoid with D$_{eq}$\,=\,1090\,km). The original density estimate of $\rho$\,=\,4.2\,\gcc\, by \citet{Fraser2010} is far outside the borders of this figure. }
    \label{fig:density}
\end{figure}

{ Our thermal emission models, which take into account the latest occultation shape and size constraints, visible light curve data (rotation period and shape constraints) from Kepler/K2, and thermal light curves at 100 and 160\,$\mu$m, provide additional constraints on Quaoar's size. While these are not our preferred solutions,} the oblate spheroid cases ({ OB1 and OB2}) provide volume-equivalent diameters of D$_{eq}$\,=\,1111\,km and D$_{eq}$\,=\,1094\,km that correspond to bulk densities of $\rho$\,=\,1.67\,\gcc\ { and $\rho$\,=\,1.75\,\gcc, respectively}, using the M$_{sys}$\,=\,1.2$\times$10$^{21}$\,kg mass estimate by \citet{Morgado2023}. Similarly, the best-fit triaxial ellipsoid case ({\sl TC}) with D$_{eq}$\,=\,1090\,km corresponds to a bulk density of $\rho$\,=\,1.77\,\gcc, and this  solution (triaxial shape) is preferred by the asymmetric double-peaked light curve and the shape differences between the 2011 and 2022 multi-chord occultations \citep{Braga-Ribas2013,Pereira2023}. Our new values above are the lowest bulk density values obtained for Quaoar so far. 

To put these densities into context, we compare them with the densities of the primaries in the most massive binary systems in the trans-Neptunian region for which reliable system mass estimates can be obtained from the satellite orbits (see Fig.~\ref{fig:density}). The current density estimates of these bodies show a strong correlation between radius and density up to a size of R$_p$\,$\lesssim$\,800\,km, roughly the size of Makemake, and a flattening for the two largest objects, Eris and Pluto. Previous estimates of Quaoar's density are clearly outside this well-defined trend (blue and purple symbols in Fig.~\ref{fig:density}). Our TPM-based estimates (orange, { pink,} and red symbols), however, fit very well to the general density-size relationship. The processes governing the formation of these large Kuiper belt objects are not fully understood \citep[see e.g.][]{BB18,Arakawa2019}. A compositionally homogeneous, rock-rich reservoir with a rock mass fraction of 70\% and a decrease in porosity with increasing size can explain density values up to $\sim$1.8\,\gcc\, \citep{BiersonNimmo2019}, but cannot account for the densities of some of the largest Kuiper belt objects (Makemake, Eris, Triton). Our density estimates suggest that the formation of Quaoar was probably not peculiar, as previous high-density values indicated, and Quaoar followed a formation mechanism similar to other large Kuiper belt objects.

\begin{acknowledgements}
PACS has been developed by a consortium of institutes led by MPE (Germany) and including UVIE (Austria); KU Leuven, CSL, IMEC (Belgium); CEA, LAM (France); MPIA (Germany); INAF-IFSI/OAA/OAP/OAT, LENS, SISSA (Italy); IAC (Spain). This development has been supported by the funding agencies BMVIT (Austria), ESA-PRODEX (Belgium), CEA/CNES (France), DLR (Germany), ASI/INAF (Italy), and CICYT/MCYT (Spain).
This work was partly supported by the K-138962 and the `SeismoLab' KKP-137523 \'Elvonal grants of the National Research, Development and Innovation Office (NKFIH, Hungary). G.M. acknowledges support by the János Bolyai Research Scholarship of the Hungarian Academy of Sciences. Part of this work was supported by the German {DLR} project number 50 OR 1108.
Part of this work was supported by the Spanish projects PID2020-112789GB-I00 from AEI and Proyecto de Excelencia de la Junta de Andalucía PY20-01309. Financial support from the grant CEX2021-001131-S funded by MCIN/AEI/ 10.13039/501100011033 is also acknowledged. 
This paper includes data collected by the Kepler mission and obtained from the MAST data archive at the Space Telescope Science Institute (STScI). Funding for the Kepler mission is provided by the NASA Science Mission Directorate.
We thank our referee for to the thorough review and the useful comments and suggestions.

\end{acknowledgements}

\bibliographystyle{aa}
\bibliography{tno}

\end{document}